\documentclass[12pt,a4paper,aps,pra,reprint,onecolumn,superscriptaddress,showpacs,nofootinbib,notitlepage,noeprint]{revtex4-2}

\usepackage{amsbsy}
\usepackage{amsfonts}
\usepackage{amsmath}
\usepackage{amssymb}
\usepackage{array}
\usepackage{bbding}
\usepackage{bbm}
\usepackage{bm}
\usepackage{curves}
\usepackage{epsfig}
\usepackage{fixmath}
\usepackage[T1]{fontenc}
\usepackage[margin=0.85in]{geometry}
\usepackage{graphicx}
\usepackage{multirow}
\usepackage{placeins}
\usepackage[svgnames]{xcolor}
\usepackage{tikz}
\usetikzlibrary{svg.path}
\usepackage[colorlinks]{hyperref}
\usepackage{relsize}
\usepackage{scalerel}

\definecolor{orcidlogocol}{HTML}{A6CE39}
\tikzset{
  orcidlogo/.pic={
    \fill[orcidlogocol] svg{M256,128c0,70.7-57.3,128-128,128C57.3,256,0,198.7,0,128C0,57.3,57.3,0,128,0C198.7,0,256,57.3,256,128z};
    \fill[white] svg{M86.3,186.2H70.9V79.1h15.4v48.4V186.2z}
                 svg{M108.9,79.1h41.6c39.6,0,57,28.3,57,53.6c0,27.5-21.5,53.6-56.8,53.6h-41.8V79.1z M124.3,172.4h24.5c34.9,0,42.9-26.5,42.9-39.7c0-21.5-13.7-39.7-43.7-39.7h-23.7V172.4z}
                 svg{M88.7,56.8c0,5.5-4.5,10.1-10.1,10.1c-5.6,0-10.1-4.6-10.1-10.1c0-5.6,4.5-10.1,10.1-10.1C84.2,46.7,88.7,51.3,88.7,56.8z};
  }
}

\newcommand\orcid[1]{\href{https://orcid.org/#1}{\mbox{\scalerel*{
\begin{tikzpicture}[yscale=-1,transform shape]
\pic{orcidlogo};
\end{tikzpicture}
}{|}}}}

\hypersetup{
	bookmarksnumbered,
	pdfstartview={FitH},
	citecolor={blue},
	linkcolor={blue},
	urlcolor={blue},
	pdfpagemode={UseOutlines}}
\definecolor{darkgreen}{RGB}{20,100,20}
\definecolor{darkblue}{RGB}{0,0,130}
\definecolor{darkred}{rgb}{.8,0,0}

\DeclareMathAlphabet{\mathantt}{OT1}{antt}{li}{it}
\DeclareMathAlphabet{\mathpzc}{OT1}{pzc}{m}{it}

\providecommand*{\I}{\mathrm{i}}                           
\newcommand{\tr}{\mathrm{tr}}													
\newcommand{\VEC}[1]{\mathbold{#1}}
\newcommand{\nn}{\nonumber}
\renewcommand{\d}{\mathrm{d}}

\newcommand{\bok}[3]{\left<\right.\hspace{-0.5ex}{#1}\left.\hspace{-0.5ex}\right|{#2}\left|\right.\hspace{-0.5ex}{#3}\left.\hspace{-0.5ex}\right>}

\newcommand{\nab}{\boldsymbol{\nabla}}

\newcommand{\e}[1]{\mathrm{e}^{#1}}

\newcommand{\Int}[1][-5pt]{\int\limits_{\begin{picture}(16,3)(-8,-3)%
		\put(0,0){\curve(-3,0,-8,0)\curve(3,0,8,0)}%
		\put(8,0){\curve(0,0,-1.5,1.5)\curve(0,0,-1.5,-1.5)}%
		\put(0,0){\arc(-3,0){180}}\put(0,0){\makebox(0,0){$\cdot$}}%
		\end{picture}}\hspace*{#1}}

\definecolor{ao(english)}{rgb}{0.0, 0.5, 0.0}

\begin{document}

\title{Phase Transitions of Repulsive Two-Component Fermi Gases in Two Dimensions}

\author{Martin-Isbj\"orn~Trappe\orcid{0000-0002-2911-4162}}
\email[]{martin.trappe@quantumlah.org}
\affiliation{Centre for Quantum Technologies, National University of Singapore, 3 Science Drive 2, Singapore 117543, Singapore}

\author{Piotr~T.~Grochowski\orcid{0000-0002-9654-4824}}
\email[]{piotr@cft.edu.pl}
\affiliation{Center for Theoretical Physics, Polish Academy of Sciences, Aleja Lotnik\'ow 32/46, 02-668 Warsaw, Poland}
 \affiliation{ICFO - Institut de Ci\`encies Fot\`oniques, The Barcelona Institute of Science and Technology, Av. Carl Friedrich Gauss 3, 08860 Castelldefels (Barcelona), Spain}

\author{Jun~Hao~Hue\orcid{0000-0003-4859-4031}}
\email[]{junhao.hue@u.nus.edu}
\affiliation{Centre for Quantum Technologies, National University of Singapore, 3 Science Drive 2, Singapore 117543, Singapore}
\affiliation{Graduate School for Integrative Sciences \& Engineering, National University of Singapore, 21 Lower Kent Ridge Road, Singapore 119077, Singapore}

\author{Tomasz~Karpiuk\orcid{0000-0001-7194-324X}}
\email[]{t.karpiuk@uwb.edu.pl}
\affiliation{Wydzia\l{} Fizyki, Uniwersytet w Bia\l{}ymstoku, ul. K. Cio\l{}kowskiego 1L, 15-245 Bia\l{}ystok, Poland}

\author{Kazimierz~Rz\k{a}\.zewski\orcid{0000-0002-6082-3565}}
\email[]{kazik@cft.edu.pl}
\affiliation{Center for Theoretical Physics, Polish Academy of Sciences, Aleja Lotnik\'ow 32/46, 02-668 Warsaw, Poland}

\date{\today}

\begin{abstract}
We predict the phase separations of two-dimensional Fermi gases with repulsive contact-type interactions between two spin components. Using density-potential functional theory with systematic semiclassical approximations, we address the long-standing problem of itinerant ferromagnetism in realistic settings. We reveal a universal transition from the paramagnetic state at small repulsive interactions towards ferromagnetic density profiles at large interaction strengths, with intricate particle-number dependent phases in between. Building on quantum Monte Carlo results for uniform systems, we benchmark our simulations against Hartree--Fock calculations for a small number of trapped fermions. We thereby demonstrate that our employed corrections to the mean-field interaction energy and especially to the Thomas--Fermi kinetic energy functional are necessary for reliably predicting properties of trapped mesoscopic Fermi gases. The density patterns of the ground state survive at low finite temperatures and confirm the Stoner-type polarization behavior across a universal interaction parameter, albeit with substantial quantitative differences that originate in the trapping potential and the quantum-corrected kinetic energy. We also uncover a zoo of metastable configurations that are energetically comparable to the ground-state density profiles and are thus likely to be observed in experiments. We argue that our density-functional approach can be easily applied to interacting multi-component Fermi gases in general.
\end{abstract}

\maketitle


\section{Introduction}
For almost a century the interacting many-body problem of quantum mechanics has been proven highly demanding both conceptually and practically.
Despite decades of intense efforts, Kohn--Sham density functional theory (DFT) \cite{Kohn1965}, the first-principles orbital-based workhorse of computational chemistry and materials science \cite{Becke2014,Hasnip2014}, remains inapplicable to large systems that are relevant in technological applications and at the forefront of fundamental research.
Especially, \textit{ab initio} descriptions of quantum gases demand a new angle of investigation.

Orbital-free DFT is the only available method for routinely and reliably computing quantum systems that harbor thousands to millions of interacting particles in nonperiodic confinement \cite{Hohenberg1964,Xia2012,Witt2018}.
Among the various flavors of orbital-free DFT, density potential functional theory (DPFT) is uniquely qualified for reliably extracting the intricate phases of interacting Fermi gases \cite{Englert1982,Englert1984,Englert1985,Englert1988,Englert1992,Trappe2016a,Trappe2017,Chau2018,Englert2019}.
DPFT reduces the many-body problem to two self-consistent equations for the single-particle density and an effective potential that includes the interaction effects.
Its capacity in simulating trapped quantum gases, especially for two-dimensional (2D) setups, extends beyond the capabilities of conventional DFT methods, which are either limited to small particle numbers \cite{Ancilotto2015,Das2018}, periodic confinement \cite{Ma2012}, or rely on ad-hoc parameterizations of the kinetic energy \cite{VanZyl2013,Gangwar2020}, although systematic gradient corrections in 2D are available for electronic systems \cite{Vilhena2014}. DPFT is a scalable approach that enables systematic semiclassical expansions beyond the Thomas--Fermi (TF) approximation across one-, two-, and three-dimensional geometries.
The conceptual, theoretical, and numerical work on DPFT over the past years has identified DPFT as an efficient, accurate, and versatile approach for targeting large-scale many-body quantum systems with arbitrary constituents, interactions, and geometries.
It has been applied to (i) noninteracting systems for benchmarking purposes \cite{Trappe2016a,Trappe2017,Chau2018,Trappe2021b}, (ii) systems in one \cite{Trappe2021a}, two \cite{Trappe2016a,Trappe2017,Trappe2019}, and three \cite{Chau2018,Trappe2021b} dimensions, (iii) small \cite{Chau2018,Trappe2021a,Trappe2021b} and large \cite{Trappe2017,Trappe2019,Trappe2021a,Trappe2021b} particle numbers, (iv) layered graphene materials \cite{Trappe2019}, (v) atomic physics \cite{Englert1982,Englert1984,Englert1985,Englert1988,Englert1992,Trappe2021b}, (vi) chemistry \cite{Trappe2021b}, and (vii) interacting Fermi gases \cite{Trappe2016a}.
The overarching feature of all these studies is the systematic methodology of DPFT, whose approximations are universally applicable to a large class of quantum systems. Our DPFT approach naturally accounts for inhomogeneities of large, trapped systems beyond the common local density approximation (LDA) and reliably yields candidates for the ground-state densities.

DPFT thus provides a natural platform to study fermion systems whose components can undergo spatial segregation due to repulsive interactions~\cite{Pethick2008,Pitaevskii2016}. One seminal example of such a behavior is itinerant ferromagnetism in metals such as iron or nickel~\cite{Giorgini2008,Brando2016}, where valence electrons spontaneously form spin-polarized domains. A quantum-mechanical description of this phenomenon has been proposed by Stoner in his mean-field model, which favors a ferromagnetic state thanks to a short-range screened Coulomb interaction that overcomes the Fermi pressure~\cite{Stoner1933}. That is, same-spin electrons congregate to form regions with nonzero net magnetization at the expense of increased kinetic energy. This simple model has fostered qualitative analyses of many-electron systems, but in other fermionic systems mechanisms beyond the short-range repulsion may suppress phase separation~\cite{Saxena2000,Pfleiderer2001}.

In this work we focus on a binary spin mixture of repulsive fermions confined to a two-dimensional harmonic trap. Even for such a rudimentary setting, the stability of a ferromagnetic (Stoner) separation is still debated both in theory~\cite{Sogo2002,Karpiuk2004,Duine2005,LeBlanc2009,Conduit2009,Cui2010,Pilati2010,Chang2011,Pekker2011,Massignan2011,Massignan2014,Levinsen2015,Trappe2016,Miyakawa2017,Koutentakis2019,Grochowski2017a,Ryszkiewicz2020,Karpiuk2020,Koutentakis2020} and in experiment~\cite{DeMarco2002,Du2008,Jo2009,Sommer2011,Sanner2012,Lee2012,Valtolina2016,Amico2018,Scazza2020}. The challenge in determining this stability stems from the competing pairing mechanism---the Feshbach resonance that is responsible for the repulsive interactions necessarily supports a weakly bound molecular state~\cite{Chin2010}. Therefore, ferromagnetic order can only manifest as an excited many-body state, in contrast to the superfluid ground state of paired fermions of opposite spins.

Experimental efforts of preparing a ferromagnetic state in an ultracold atomic system of a balanced mixture of the two lowest hyperfine states of lithium-6 date back to the late 2000s. The initial attempts of observing a para- to ferromagnetic transition proved inconclusive, though some signatures, such as an increase of the kinetic energy, supported its existence~\cite{Jo2009,Sanner2012}. The ambiguity came from an alternative explanation of the rapid molecule formation that could produce similar results.
To circumvent this problem, the system was prepared in an artificial domain structure, where each of the components initially reside in their respective half of the harmonic trap~\cite{Sommer2011,Valtolina2016}. Such a setup showed stability for a finite time, which was later confirmed by more advanced time-resolved studies of the competition between pairing and ferromagnetic instabilities~\cite{Amico2018,Scazza2020}.

The theoretical treatment, on the other hand, has been continuously refined in recent history. The analysis of the purely repulsive ground state of a balanced system has been studied with many different approaches, based on second-order perturbation theory~\cite{Duine2005}, Landau's Fermi liquid theory with state-of-the-art Quantum Monte Carlo simulations~\cite{Conduit2009,Pilati2010,Chang2011}, lowest-order constraint variational calculation~\cite{Heiselberg2011}, nonperturbative ladder approximation ~\cite{He2012}, large-$N$ expansion, dimensional $\epsilon$-expansion~\cite{He2016}, and pseudo-Schr\"odinger evolution~\cite{Trappe2016,Grochowski2017a}. In a three-dimensional (3D) geometry, each of these methods suggested the existence of the ferromagnetic transition, though with a varying critical value of interaction strength, depending on the approach that was used.

The 2D setup considered here potentially offers an escape from the stability problem, as three-body recombination processes are less important in lower dimensions~\cite{Giorgini2008}. However, the pairing mechanisms in 2D differ from their 3D counterparts and may still preclude a stable phase-separated state~\cite{Levinsen2015}. Moreover, experimental data suggests that in an impurity limit, relaxation to the bound state plays a crucial role~\cite{Koschorreck2012}. Also theoretical approaches, including mean-field, perturbative and diagrammatic expansions~\cite{Bloom1975,He2014}, polaronic approach~\cite{Schmidt2012,Ngampruetikorn2012}, and quantum Monte Carlo methods~\cite{Conduit2010,Conduit2013,Bertaina2013,Whitehead2016}, have not unambiguously predicted the Stoner transition, not even for purely repulsive mixtures. The subtle interplay between interaction and kinetic energies, an essence of the Stoner ferromagnetism, is greatly affected by quantum correlations and as such, evaluation of beyond-mean-field effects is crucial for such investigations. Also other types of interactions such as Rabi coupling~\cite{Penna2017} and dipolar forces~\cite{Comparin2019} have been analyzed recently in search of stable ferromagnetic phases in 2D ultracold gases. 

In this work, we aim at an unambiguous and quantitatively reliable picture of the phase transitions of a repulsive balanced mixture in realistic settings, that is, for large particle numbers and an inhomogeneous trapping potential. We base our predictions on two different approaches of kinetic energy evaluation, which are discussed in section~\ref{kinetic}. One is based on DPFT, which we introduce in section~\ref{SecDPFT} and systematically approximate in section~\ref{SectionST}; see also appendix~\ref{AppendixDPFT}. The other is the multi-component Hartree-Fock (HF) method, see section~\ref{SecHF} and appendix~\ref{AppendixHF}. One option for assessing the robustness of a phase transition in realistic settings is to vary the interaction energy functional. section~\ref{interaction} specifies how we utilize two different interactions---a bare contact, viz., mean-field repulsion and a quantum Monte Carlo energy functional in local density approximation, which we term `renormalized' contact interaction henceforth. We perform a thorough analysis of emerging partially separated density profiles and their dependence on either of the two energy functionals. We present our main results in section~\ref{Results}. By comparing with DPFT simulations for the mean-field contact interaction in section~\ref{SectionTF}, we elucidate the inadequacy of the TF model for describing multi-component Fermi gases with contact-type interactions. In section~\ref{DPFTvsHF}, we benchmark the consequently required quantum corrections beyond the TF approximation against HF results for both interaction energy functionals. This enables us in sections~\ref{RenCon} and \ref{MetastableDensities} to (i) reliably predict semiclassical DPFT density profiles of both ground- and metastable states for large particle numbers across interaction strengths and (ii) compare the polarization curve of the resulting phase transition against the Stoner-type QMC prediction for uniform systems. Section~\ref{conclusions} summarizes our findings and points at further, potentially fruitful continuations of our work.

\FloatBarrier

\section{Kinetic energy functionals beyond the Thomas--Fermi approximation} \label{kinetic}

\subsection{\label{SecDPFT}Multi-component density-potential functional theory}
A prerequisite for accurate densities and energies is a sufficiently accurate kinetic energy. Ideally, one employs the computationally most efficient TF approximation of the kinetic energy, which is adequate for selected systems. However, the TF density $n_{\mathrm{TF}}$ often cannot even qualitatively describe the physics, for example, of a contact-interacting two-component Fermi gas \cite{Trappe2016}; see also figure~\ref{TFvsn3p} in appendix~\ref{AppendixDPFT}. In any case, quantum-corrected density formulae have to either validate or replace $n_{\mathrm{TF}}$. While von-Weizs\"acker-type gradient corrections to the TF approximation of the kinetic energy density functional are successfully used for three-dimensional geometries \cite{Dreizler1990,Garcia-Aldea2012}, also in the context of ultracold Fermi gases \cite{Trappe2016,Miyakawa2020}, attempts of systematically deriving its 2D analog have produced ambiguous results at best \cite{Holas1991,Shao1993,VanZyl2001,Brack2003,Salasnich2007,Koivisto2007,Putaja2012}. Density potential functional theory presents an unambiguous solution to this dilemma \cite{Trappe2016a,Trappe2017}. What sets DPFT apart from other orbital-free DFT approaches are systematic quantum corrections to the TF approximation, which are not available for the commonly used \textit{density} functional $E_{\mathrm{kin}}[n]$ of the kinetic energy in 2D. We are equipped with two such approximation schemes that rely on the explicitly available expression of the Legendre transform of $E_{\mathrm{kin}}[n]$, the \textit{potential} functional $E_1[V]$, which is expressed as a single-particle trace that can be systematically approximated using semiclassical techniques. One scheme delivers nonlocal density formulae from a split-operator approximation of the quantum-mechanical propagator \cite{Chau2018,Trappe2019,Trappe2021a,Trappe2021b}. The other scheme is based on the Wigner function formalism and Airy-averaging techniques \cite{Englert1988,Trappe2016a,Trappe2017,Trappe2021a,Trappe2021b}. Both schemes are well-established with a track-record of excellent accuracy and computational efficiency for a large variety of systems from harmonium to Fermi gases to electron-hole distributions in layered materials \cite{Englert1988,Trappe2016a,Trappe2017,Chau2018,Trappe2019,Trappe2021a,Trappe2021b}. However, in this work we shall focus on the first scheme, since the semilocal `Airy-averaged' densities inherit some of the shortcomings of the inadequate TF model when applied to multi-component contact-interacting systems; see appendix~\ref{AppendixDPFT} for further details.

In this section, we present the straightforward multi-species extension of the DPFT formalism. In any orbital-free DFT approach, the stationary points of the constrained density functional of the total energy 
\begin{align}\label{gsEnergy2}
E&=E[\VEC n,\VEC \mu](\VEC N)=E_{\mathrm{kin}}[\VEC n]+E_{\mathrm{ext}}[\VEC n]+E_{\mathrm{int}}[\VEC n]+\VEC\mu\cdot\left(\VEC N-\int(\d\VEC r)\,\VEC n(\VEC r)\right)
\end{align}
deliver the ground-state densities ${\VEC n(\VEC r)=\{n_1(\VEC r),\dots,n_S(\VEC r)\}}$ that integrate to the chosen particle numbers ${\VEC N=\{N_1,\dots,N_S\}}$ for $S$ species of particles (enforced through the chemical potentials ${\VEC \mu=\{\mu_1,\dots,\mu_S\}}$), thereby producing the proper trade-off between kinetic ($E_{\mathrm{kin}}$), external ($E_{\mathrm{ext}}$), and interaction energy ($E_{\mathrm{int}}$). In DPFT, we introduce an auxiliary variable, the effective potential energy
\begin{align}\label{deltaEkin}
V_s(\VEC r)=\mu_s-\frac{\delta E_{\mathrm{kin}}[n_s]}{\delta n_s(\VEC r)}
\end{align}
for species $s\in\{1,\dots,S\}$, such that the Legendre transform
\begin{align}\label{LegendreTF}
  E_1[\VEC V-\VEC \mu]=E_{\mathrm{kin}}[\VEC n]
             +\int(\d\VEC r)\,\big(\VEC V(\VEC r)-\VEC \mu\big)\cdot\VEC n(\VEC r)
\end{align}
of the kinetic energy functional
\begin{align}\label{Ekin}
E_{\mathrm{kin}}[\VEC n]=\sum_sE_{\mathrm{kin}}[n_s]
\end{align}
transforms equation~(\ref{gsEnergy2}) into
\begin{align}\label{EnergyVnmu}
E&=E[\VEC V,\VEC n,\VEC \mu](\VEC N)=E_1[\VEC V-\VEC \mu]-\int(\d\VEC r)\,\VEC n(\VEC r)\cdot\big(\VEC V(\VEC r)-\VEC V^{\mathrm{ext}}(\VEC r)\big)+E_{\mathrm{int}}[\VEC n]+\VEC\mu\cdot\VEC N\,.
\end{align}
Strictly, equation~(\ref{Ekin}) holds only for independent particles, but it can be made exact by transferring the interacting part of the kinetic energy into the interaction energy functional $E_{\mathrm{int}}[\VEC n]$. For each species $s$, the $V_s$- and $n_s$-variations at the stationary points of $E[\VEC V,\VEC n,\VEC \mu]$ obey
\begin{align}
  \delta V_s: & \qquad \,n_s[V_s-\mu_s](\VEC r)=
              \frac{\delta E_1[V_s-\mu_s]}{\delta V_s(\VEC r)}\label{n}
\intertext{and}
 \delta n_s: & \qquad\label{V}
          V_s[\VEC n](\VEC r)=V_s^{\mathrm{ext}}(\VEC r)
          +\frac{\delta E_{\mathrm{int}}[\VEC n]}{\delta n_s(\VEC r)}\,,
\end{align}
respectively. The $\mu_s$-variation, combined with equation~(\ref{n}), reproduces the particle-number constraint  
\begin{align}\label{PartNumConstraint}
\int(\d\VEC r)\,n_s(\VEC r)=N_s\,.
\end{align}
Equation~(\ref{n}) states that the particle density is a functional of the effective potential and immediately yields the particle density in the noninteracting case (${V_s=V_s^{\mathrm{ext}}}$) for any given $\mu_s$. Conversely, equation~(\ref{V}) declares $V_s$ as a functional of all densities $\VEC n$, such that a self-consistent solution of equations~(\ref{n})--(\ref{PartNumConstraint}) for any interaction functional $E_{\mathrm{int}}[\VEC n]$ produces the ground-state density, much like in the Kohn--Sham scheme, but without resorting to orbitals. We initialize the self-consistent loop with ${n_s^{(0)}=n_s\left[V_s^{(0)}=V_s^{\mathrm{ext}}\right]}$ and iterate the densities via
\begin{align}\label{SCloop}
n_s^{(i)}&\overset{\mbox{\footnotesize~(\ref{V})}}{\xrightarrow{\hspace*{3em}}}V_s^{(i+1)}=V_s\left[\VEC n^{(i)}\right]\overset{\mbox{\footnotesize~(\ref{n})}}{\xrightarrow{\hspace*{3em}}}n_s^{(i+1)}=(1-\theta_s)\,n_s^{(i)}+\theta_s\,n_s\left[V_s^{(i+1)}-\mu_s^{(i+1)}\right],
\end{align}
until all densities $\VEC n$ have converged with the help of mixing parameters ${\VEC\theta=\{\theta_1,\dots,\theta_S\}}$. The chemical potentials $\VEC\mu$ are adjusted in each iteration to enforce the particle number constraints of equations~(\ref{PartNumConstraint}). Further details of the numerical implementation of equation~(\ref{SCloop}) and implications of equations~(\ref{EnergyVnmu})--(\ref{PartNumConstraint}) are well documented in the literature, see references~\cite{Englert1988,Trappe2016a,Trappe2017,Englert1992,Trappe2019} and references therein.

The key element of DPFT is the \textit{potential} functional ${E_1[\VEC V-\VEC \mu]}$, which captures the effects of the kinetic energy in place of the \textit{density} functional $E_{\mathrm{kin}}[n]$. The explicit form of $E_{\mathrm{kin}}[n]$ is unknown even for noninteracting systems and its approximations can compete with Kohn--Sham computations only for selected systems. In contrast, $E_1[\VEC V-\VEC \mu]$ is explicitly available for independent particles \cite{Englert1992} in terms of single-particle traces, which can be approximated systematically:
\begin{align}\label{tracef}
E_1[\VEC V-\VEC\mu]=\sum_{s=1}^S E_1[V_s-\mu_s]=\sum_{s=1}^S\tr\left\{\mathcal E_T^{(0)}(H_s-\mu_s)\right\}
\end{align}
for temperatures ${T\ge0}$, with the function
\begin{align}\label{fHmuT}
  \mathcal E_T^{(0)}(A_s=H_s-\mu_s)=(-k_{\mathrm{B}}T)\,\log{\left(1+\mathrm{e}^{-A_s/k_{\mathrm{B}}T}\right)}
\end{align}
of the single-particle Hamiltonian
\begin{align}\label{singlePartHamil}
H_s(\VEC R,\VEC P)=\frac{\VEC P^2}{2m}+V_s(\VEC R).
\end{align}
The single-particle position and momentum operators for $D$ Cartesian dimensions are $\VEC R$ and $\VEC P$, respectively. Here and in the following we omit arguments of functions for brevity wherever the command of clarity permits.
Although the (unknown) interacting part $\Delta E_{\mathrm{kin}}$ of the kinetic energy can formally be transferred into the interaction energy such that equation~(\ref{tracef}) becomes exact as part of the total energy in equation~(\ref{EnergyVnmu}), we neglect $\Delta E_{\mathrm{kin}}$ altogether---a procedure that comes with an excellent track-record also for DPFT \cite{Englert1988,Trappe2016a,Trappe2017,Chau2018,Trappe2019,Trappe2021a,Trappe2021b} and reiterates the fact that $\Delta E_{\mathrm{kin}}$ is often of secondary importance. In the limit of zero temperature we obtain the ground-state version
\begin{align}
\mathcal E_0^{(0)}(H_s-\mu_s)=(H_s-\mu_s)\,\Theta(\mu_s-H_s)
\end{align}
of equation~(\ref{fHmuT}); the step function $\Theta(\,)$ is the zero-temperature limit of ${\Theta_T(\mu_s-H_s)=\left[1+\mathrm{e}^{(H_s-\mu_s)/k_{\mathrm{B}}T}\right]^{-1}}$. Any approximation of the single-particle trace in equation~(\ref{tracef}) yields an according approximation for the particle density in equation~(\ref{n}). We can benchmark semiclassical approximations of $E_1$ unambiguously if the interaction functional is known exactly or for any noninteracting system, as done in references~\cite{Trappe2016a,Trappe2017,Chau2018,Trappe2021a,Trappe2021b}.

\subsection{\label{SectionST}Densities and energies from Suzuki--Trotter-factorized time-evolution operator}

Our approximation schemes for DPFT are based on semiclassical expansions of the trace in equation~(\ref{tracef}), which includes a degeneracy factor $g$ (e.g., spin-multiplicity). We reiterate some of the results in~\cite{Chau2018} and begin with realizing that equations~(\ref{n}) and (\ref{tracef}) at zero temperature yield
\begin{align}\label{nSTA}
n(\VEC r)=g\bok{\VEC r}{\Theta(\mu-H)}{\VEC r}
         =g\Int\frac{\d t}{2\pi\I t}\,\e{\frac{\I t}{\hbar}\mu}\,
           \bok{\VEC r}{U(t)}{\VEC r},
\end{align}
which invites tailored Suzuki--Trotter (ST) factorizations of the unitary time-evolution operator ${U(t)=\e{-\frac{\I t}{\hbar}H}}$. In equation~(\ref{nSTA}), we make use of the Fourier transform of the step function $\Theta(\ )$, and the integration path from ${t=-\infty}$ to ${t=\infty}$ crosses the imaginary $t$ axis in the lower half-plane. For notational brevity, we drop the subscript $s$ in all species-dependent variables (the here developed formulae hold for all $s$ individually). We obtain a hierarchy of approximations of $n(\VEC r)$ from appropriate coefficients $a_k$ and $b_k$ of the ansatz
\begin{align}\label{Unu}
  U(t)\approx U_\nu(t)
  =\prod_{k=1}^{\lceil \nu/2 \rceil} \e{- \frac{\I t}{\hbar} a_k
    V(\VEC{R})}\, \e{- \frac{\I t}{\hbar} b_k \VEC{P}^2/(2m)},
\end{align}
where the exponential factors are multiplied from left to right in order of increasing $k$-values. Equation~(\ref{Unu}) creates a series of increasingly accurate semiclassical approximations beyond the TF approximation without a gradient expansion.

Reference \cite{Chau2018} reports particle densities $n_\nu$ based on up to $\nu=7$ exponential factors. The (generically) most accurate approximation $U_7$ in~\cite{Chau2018} has been proven highly accurate for a variety of systems \cite{Chin1997,Omelyan2002,Chin2005,Chau2018,Hue2020}. However, the computational cost of $n_\nu(\VEC r)$ for ${\nu>3}$ in the currently available spatially explicit formulations can be prohibitive at high spatial resolutions and in the case of slow convergence of equation~(\ref{SCloop}). We therefore focus on a three-factor approximation, which has been successfully used in layered 2D materials \cite{Trappe2019}. Choosing ${a_1=0,\;a_2=1,\;b_1=b_2=1/2}$, we obtain
\begin{align}\label{n3p}
n_{3'}(\VEC r)=g\int(\d\VEC r')\left(\frac{k_{3'}}{2\pi r'}\right)^D J_D(2r'\,k_{3'}),
\end{align}
with the Bessel function $J_D(\,)$ of order $D$ and the effective Fermi wavenumber
\begin{align}
k_{3'}=\frac{1}{\hbar}\big[2m\big(\mu-V(\VEC r+\VEC r')\big)\big]_+^{1/2},
\end{align}
where ${[z]_+=z\,\Theta(z)}$. The approximate density formula $n_{3'}(\VEC r)$ is the quantum-corrected successor of the TF density
\begin{align}\label{nTF}
n_{\mathrm{TF}}(\VEC r)=n_2(\VEC r)&=\frac{g\,\Omega_D}{D\,(2\pi\hbar)^D}\big[2m\,\big(\mu-V(\VEC r)\big)\big]_+^{D/2}=\frac{g\,\Omega_D}{D\,(2\pi\mathcal{U}^2)^D}\big[2\,\big(\mu-V(\VEC r)\big)\big]_+^{D/2},
\end{align}
with solid angle $\Omega_D$ in $D$ dimensions, which is obtained from the two-factor splitting ${a_1=b_1=1}$ that neglects the noncommutativity of $\VEC R$ and $\VEC P$.
In all formulae of this work that exhibit the dimensionless constant
\begin{align}
\mathcal{U}=\hbar^2/(m\,\mathcal L^2\,\mathcal E),
\end{align}
the quantities of energy are given in units of $\mathcal E$ and those of length in units of $\mathcal L$. For example, $\mu$ in equation~(\ref{nTF}) is implicit for $\mu/\mathcal E$, and $n_{\mathrm{TF}}(\VEC r)$ comes in units of $\mathcal L^{-D}$, which are not explicitly exhibited in equation~(\ref{nTF}). For the concrete examples in the sections below we use harmonic oscillator units ${\mathcal E=\hbar\omega}$ and ${\mathcal L=\sqrt{\hbar/(m\,\omega)}}$, which imply ${\mathcal{U}=1}$ and ${V^{\mathrm{ext}}(\VEC r)=\frac{\mathcal E}{2}(\VEC r/\mathcal L)^2}$.

In contrast to the local TF density, whose computational cost scales with size $G$ of the numerical grid, $n_{3'}(\VEC r)$ is a fully nonlocal expression, which samples the effective potential $V$ in a neighborhood of the position $\VEC r$. This effective averaging over whole regions of position space provides a quantum correction to the TF density and allows for density tails in the classically forbidden region of $V$, but also lets the computational cost of $n_{3'}$ scale like $G^2$. In this work we make extensive use of the zero-temperature expression $n_{3'}$ from equation~(\ref{n3p}), which is efficient enough for isotropic calculations. For anisotropic calculations at $T=0$, it is expedient to rephrase $n_{3'}$ in terms of Fast Fourier Transforms. We derive the according expression $n_{3'}^{\mathcal F}$ as follows. Upon approximating the time-evolution operator by $U_{3'}$ and inverting the Fourier transform of $\Theta(\,)$ in equation~(\ref{nSTA}), we arrive at
\begin{align}\label{nSTAT}
n_{3'}(\VEC r)=g\int\frac{(\d\VEC p_1)(\d\VEC p_2)(\d\VEC r_1)}{(2\pi\hbar)^{2D}}\,\exp\left(\frac{\I}{\hbar}\VEC r_1\cdot(\VEC p_1-\VEC p_2)\right)\,\Theta(\mu-H_{3'}),
\end{align}
where ${H_{3'}=\frac{\VEC p_1^2+\VEC p_2^2}{4m}+V(\VEC r+\VEC r_1)}$. With ${\VEC p=\hbar\VEC k}$ we write\footnote{We denote the Fourier transform of a function $f(\VEC r)$ as ${f(\VEC k)=\mathcal F\{f(\VEC r)\}(\VEC k)=\int(\d\VEC r)\,\e{-\I\VEC k\VEC r}\,f(\VEC r)}$ and implement $\mathcal F$ as a Fast Fourier Transform using the FFTW library for C++ \cite{Frigo2005}.}
\begin{align}
n_{3'}(\VEC k)&=\mathcal F\{n_{3'}(\VEC r')\}(\VEC k)=\int(\d\VEC r')\,\e{-\I\VEC k\VEC r'}\,n_{3'}(\VEC r')\nn\\
&=\frac{g}{(2\pi)^{2D}}\int(\d\VEC k_1)(\d\VEC k_2)(\d\VEC r_1)\,\e{\I\,\VEC r_1\cdot(\VEC k_1-\VEC k_2)}\int(\d\VEC r')\,\e{-\I\VEC k\VEC r'}\Theta\left(\mu-\frac{\VEC k_1^2+\VEC k_2^2}{4m/\hbar^2}-V(\VEC r'+\VEC r_1)\right)\label{n3pTksecond}
\end{align}
for the Fourier transform of $n_{3'}(\VEC r)$. Defining ${\VEC r_2=\VEC r'+\VEC r_1}$, we express the last integral in equation~(\ref{n3pTksecond}) as 
\begin{align}
\e{\I\VEC k\VEC r_1}\int(\d\VEC r_2)\,\e{-\I\VEC k\VEC r_2}\Theta\left(\mu-\frac{\VEC k_1^2+\VEC k_2^2}{4m/\hbar^2}-V(\VEC r_2)\right), 
\end{align}
such that
\begin{align}\label{n3pTappendix}
n_{3'}(\VEC k)&=\frac{g}{(2\pi)^{D}}\int(\d\VEC k_1)(\d\VEC k_2)\underset{\delta(\VEC k+\VEC k_1-\VEC k_2)}{\underbrace{\int\frac{(\d\VEC r_1)}{(2\pi)^{D}}\,\e{\I\,\VEC r_1\cdot(\VEC k+\VEC k_1-\VEC k_2)}}}\times\int(\d\VEC r')\,\e{-\I\VEC k\VEC r'}\Theta\left(\mu-\frac{\VEC k_1^2+\VEC k_2^2}{4m/\hbar^2}-V(\VEC r')\right)\\
&=\frac{g\,\Omega_D}{(2\pi)^{D}}\int(\d\VEC r')\,\e{-\I\VEC k\VEC r'}\int_0^\infty\d k_1\,k_1^{D-1}\,\Theta\left(Q^2-\hbar^2\VEC k_1^2\right)=\frac{g\,\Omega_D}{(2\pi)^{D}}\int(\d\VEC r')\,\e{-\I\VEC k\VEC r'}\frac{Q^D}{D}\Theta\left(Q^2\right),
\end{align}
where ${Q^2=2m\big(\mu-V(\VEC r')\big)-\frac{\hbar^2\VEC k^2}{4}}$. We then write 
\begin{align}\label{n3pF}
n_{3'}^{\mathcal F}(\VEC r)=\frac{g\,\Omega_D}{D\,(2\pi)^{D}}\mathcal F^{-1}\left\{\int(\d\VEC r')\,\e{-\I\VEC k\VEC r'}\,\left[\frac{2}{\mathcal{U}}\big(\mu-V(\VEC r')\big)-\frac{\VEC k^2}{4}\right]_+^{D/2}\right\}(\VEC r)
\end{align}
for $n_{3'}(\VEC r)$ to distinguish equation~(\ref{n3pF}) from the (numerically identical) expression of equation~(\ref{n3p}). The computational cost of $n_{3'}^{\mathcal F}$ still scales like $G^2$, but is reduced by a factor of $\sim$10--40 since only exponentials (not Bessel functions) have to be evaluated.

The kinetic energy $E_{\mathrm{kin}}$ is a functional of $n$, not of $V$, but its $U_{3'}$-approximated stationary value
\begin{align}\label{Ekin3p}
E_{\mathrm{kin}}^{(3')}=\frac{g\,\Omega_D}{(2\pi\mathcal{U}^2)^D\,(2D+4)}\int(\d\VEC r)\,\big[2\big(\mu-V(\VEC r)\big)\big]_+^{\frac{D+2}{2}}
\end{align}
can be calculated from the ground-state $V$, see appendix~\ref{AppendixDPFT}. Incidentally, equation~(\ref{Ekin3p}) coincides with the TF kinetic energy $E_{\mathrm{kin}}^{\mathrm{TF}}$, whose ground-state values (evaluated with the ground-state effective potential) equal those of the density functional 
\begin{align}
E_{\mathrm{kin}}^{\mathrm{TF}}[n]=\frac{2\pi\,\mathcal{U}\,D^{(D+2)/D}}{(g\,\Omega_D)^{2/D}\,(2D+4)}\int\d\VEC r\,n(\VEC r)^{\frac{D+2}{2}}
\end{align}
(evaluated with the ground-state density) upon translating\footnote{Note that the TF approximation in equation~(\ref{nTF}) allows for point-wise identification of $n$ and $V$ only in the classically allowed region.} between $n$ and $V$ via equation~(\ref{nTF}). Quantitative differences between ${E_{\mathrm{kin}}^{(3')}[V-\mu]}$ and ${E_{\mathrm{kin}}^{\mathrm{TF}}[V-\mu]}$ at the stationary point of the total energy are generally expected since ${n_{3'}\not=n_{\mathrm{TF}}}$ and ${V_{3'}\not=V_{\mathrm{TF}}}$ at the ground state.

\subsection{\label{SecHF}Hartree--Fock approach to trapped multi-component fermion gases}

In this work, we benchmark $n_{3'}$-based DPFT densities against Hartree--Fock (HF) results. The derivation of the HF equations
\begin{eqnarray}\label{HFeqweuse}
 \left[ -\frac{\hbar^2}{2 m} \nabla^2 + V_{\text{ext}}(\VEC r)
+ \frac{\delta E_{\text{int}}[\VEC n]}{\delta n_{s} (\VEC r)} \; \right] \; \varphi_i^{(s)} (\VEC r) & = & \varepsilon_i^{(s)} \, \varphi_i^{(s)}(\VEC r)
\end{eqnarray}
for the two-component spin mixture (${s\in\{1,2\}}$) is shown in appendix~\ref{AppendixHF}. Here, $\varphi_i^{(1)}(\VEC r)$ and $\varphi_i^{(2)}(\VEC r)$, with ${i=1,...,N/2}$, are spin-orbitals of the first and the second spin component, respectively. We deploy the interaction terms $\frac{\delta E_{\text{int}}[\VEC n]}{\delta n_{1/2} (\VEC r)}$ as defined below through equations~(\ref{EintPureCon}) and (\ref{EintRenCon}) for the mean-field and the renormalized contact interaction, respectively. The one-particle densities
\begin{eqnarray}
 n_s(\VEC r) & = & \sum_{i=1}^{N/2} |\varphi_i^{(s)} (\VEC r)|^2
\end{eqnarray}
associated with the spin components $s$ sum to the total one-particle density $n(\VEC r) = n_1(\VEC r)+n_2(\VEC r)$.

\section{Interaction energy of a two-component repulsive Fermi gas} \label{interaction}
Many studies have shown that the simple mean field interaction functional
\begin{align}\label{EintPureCon}
E_{\mathrm{int}}^{\mathrm{MF}}[\VEC n=(n_1,n_2)]=\int(\d\VEC r)\,\alpha\,n_1(\VEC r)\,n_2(\VEC r)
\end{align}
needs to be renormalized in order to reproduce experimental data~\cite{Levinsen2015}. Different interaction regimes of homogeneous two-dimensional Fermi gases are commonly defined through the dimensionless gas parameter
\begin{align}\label{eta}
\eta = -1/\log(k_{\mathrm{F}} a_{\text{2D}}), 
\end{align}
where ${k_{\mathrm{F}} = \sqrt{{2 \pi n}}}$ is the Fermi wavenumber, and $a_{\text{2D}}$ is the scattering length in 2D. Out of the two common definitions of $a_{\text{2D}}$ that are based on the two-body scattering problem, we use the one in which the energy of the two-body bound state equals ${\epsilon_b = -4 \hbar^2 /\left(  m\, a_{\text{2D}}^2\,e^{2 \gamma}\right)} $, where ${\gamma \approx 0.577}$ is the Euler--Mascheroni constant.

The many-body ground state of a two-component ultracold Fermi gas is called the lower (attractive) branch of the energy spectrum. For this lower branch, the crossover from the BEC regime of tightly bound dimers (${\eta>0}$ and ${\eta \ll 1}$) to the BCS superfluid regime ($\eta<0$ and $\eta \ll - 1$) has been thoroughly investigated.

The upper (repulsive) branch refers to the excited state of the many-body spectrum that exhibits repulsive behavior.
It is usually associated with the para- to ferromagnetic transition of the Fermi gas, above which the polarized mixture is favored energetically over the spin-balanced one.
Such a `Stoner' phase transition is named after Edmund Stoner who devised an early mean-field description of ferromagnetism in the 1930s.
Stoner ferromagnetism has been studied extensively, both theoretically and experimentally, mostly in 3D settings.
However, the decay into the energetically more favorable lower branch is the main obstacle towards the experimental realization of phase-separated states and invites the precise study of competition between pairing and anticorrelating dynamics. 
However, this issue can be resolved by introducing an artificial domain structure or through a fast interaction quench.

To date, theoretical studies have been focusing mainly on the homogeneous mixture. The usual approach for describing the homogeneous system is based on the Jastrow--Slater ansatz for the many-body wave function that includes only two-body correlations. 
With this ansatz, the energy can be minimized, for example, by a Quantum Monte Carlo (QMC) scheme or by introducing constraints as in the lowest-order constrained variational (LOCV) approximation.
While QMC is believed to be the most accurate method available, popular alternatives include perturbative methods, either through a diagrammatic expansion, a polaron approach or a large-$N$ expansion.

\begin{figure}[htb!]
\includegraphics[width=0.35\linewidth]{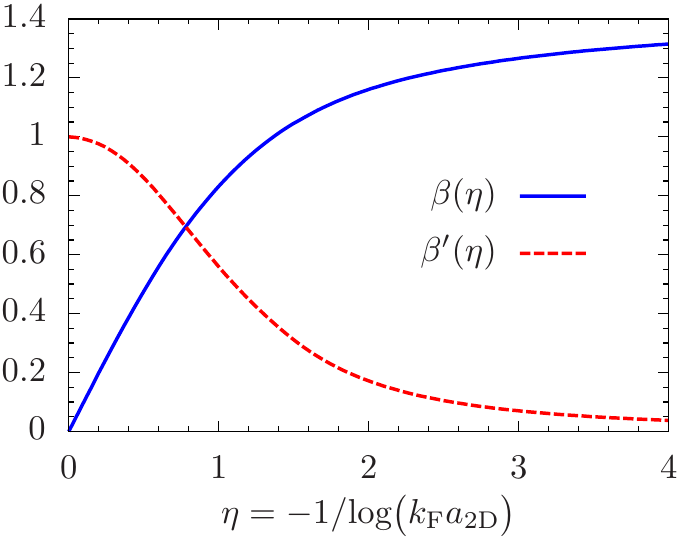}
\includegraphics[width=0.375\linewidth]{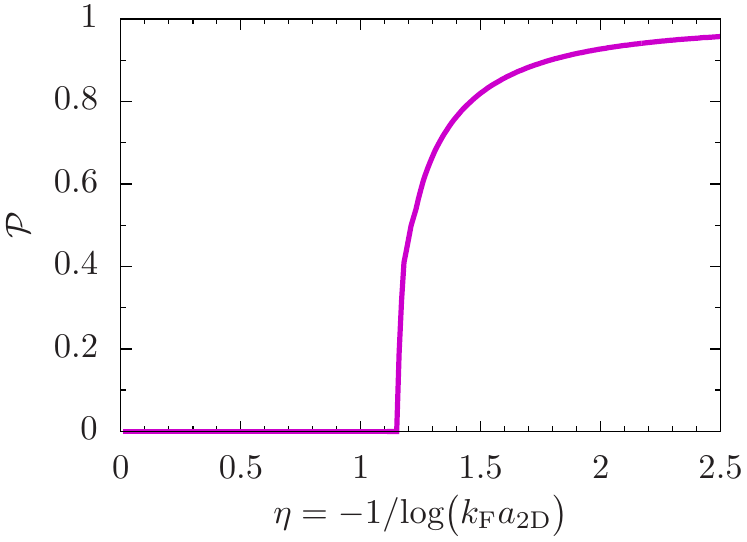}
\caption{\label{betaM} Left: The ratio $\beta$ between the interaction energy and the kinetic energy of the noninteracting system and its derivative as a function of the dimensionless parameter $\eta$.
We evaluate $\beta$ through a quantum Monte Carlo approach with the help of the CASINO package, using a smooth \textit{ultratransferable} potential~\cite{Whitehead2016} as an inter-particle interaction.
Right: The ground-state polarization of the uniform two-component Fermi mixture. The sample becomes partially polarized at a critical interaction strength of $\eta \approx 1.15$.}
\end{figure}

In this work, we parameterize the interaction energy functional using QMC results.
For a homogeneous system, it is expedient to introduce the ratio between the interaction energy and the kinetic energy $E_{\text{kin}}^0$ of the noninteracting system:
\begin{align}
    \beta(\eta) = \frac{E_{\text{int}}}{E_{\text{kin}}^0} = \frac{E_{\text{tot}}-E_{\text{kin}}^0}{E_{\text{kin}}^0},
\end{align}
which is expressed in terms of the dimensional gas parameter $\eta$.
To compute $\beta$, we use the CASINO package with a smooth pseudopotential devised by Whitehead \textit{et al.}~\cite{Whitehead2016} that captures the effective interactions between fermions from different species.
In choosing the closed-shell structure of ${49+49}$ fermions, we reduce finite-size effects.
We approximate the resulting parameterization by a polynomial
\begin{align}
\beta(\eta)=
\begin{cases}
 -0.061398\, \eta ^6+0.25332\, \eta ^5-0.30739\, \eta ^4-0.058454\, \eta ^2+1.0062\, \eta -0.00041475 & ,\;0\leq\eta\leq 1.55 \\
 1.4436 -0.46338/\eta -0.20465/\eta ^2 & ,\;\eta >1.55
\end{cases} ,
\end{align}
which is depicted in figure~\ref{betaM} (left), together with its derivative $\beta'(\eta)$.
Note that we restrict ourselves to positive values of $\eta$, which corresponds to ${k_{\mathrm{F}} a_{\text{2D}}<1}$ and ensures that the mixture stays on the repulsive branch.

Since we address inhomogeneous systems using a local density approximation, we introduce local Fermi wavenumbers ${k_{\text{F}}^s=\sqrt{4 \pi n_s}}$ (here, ${s\in\{1,2\}}$) and local gas parameters ${\eta_s = - 1 / \log \left(k_{\text{F}}^s  a_{\text{2D}} \right)}$ as direct extensions of their homogeneous versions.
In order to consider locally spin-imbalanced systems, we perform a simple symmetrization to obtain the total energy,
\begin{align}\label{EtotHomo}
E_{\text{tot}} = \sum_s \frac{1}{2} n_{s} \epsilon_F^{s} + \sum_{\overset{\scriptstyle{s,s'}}{s\not=s'}} \frac{1}{2} n_{s} \epsilon_F^{s '} \beta(\eta_{s'}),
\end{align}
consisting of kinetic and interaction terms  with local Fermi energies ${\epsilon_F^s = 2 \hbar^2 \pi n_s / m}$. 
In the spirit of the local density approximation, equation~(\ref{EtotHomo}) invites us to address heterogeneous setups using the approximate density functional
\begin{align}\label{EintRenCon}
E_{\text{int}}[\VEC n]=\int(\d\VEC r)\,\epsilon_{\text{int}}[\VEC n](\VEC r) 
\end{align}
of the interaction energy, with
\begin{align}\label{epsilonint}
\epsilon_{\text{int}}[\VEC n](\VEC r) = \frac{\pi \hbar^2}{m} n_1 (\VEC r) n_2 (\VEC r) \big[ \beta\big(\eta_{1} (\VEC r)\big) + \beta\big(\eta_{2} (\VEC r)\big)\big]
\end{align}
and
\begin{align}\label{epsilonintderiv}
\frac{\delta E_{\text{int}}[\VEC n]}{\delta n_{1/2} (\VEC r)} = \frac{\pi \hbar^2}{m} n_{2/1} (\VEC r) \left[ \beta\big(\eta_{1} (\VEC r)\big) + \beta\big(\eta_{2} (\VEC r)\big) + \frac{1}{2} \eta^2_{1/2} \beta'\big(\eta_{1/2} (\VEC r)\big)\right].
\end{align}

Before studying inhomogeneous systems, let us analyze a homogeneous one with a fixed density ${n=n_1 + n_2}$.
Defining relative densities ${x = n_1 / n}$ and ${1-x = n_2 / n}$, we rewrite the total energy as 
\begin{align}\label{EtotStoner}
E_{\text{tot}} =& \frac{n^2 \pi \hbar^2}{m} \left[ 1 + x (1-x) \left( \beta \left( -\frac{1}{\log{ \sqrt{2 x}}-1/\eta}  \right) + \beta \left( -\frac{1}{\log{ \sqrt{2(1- x)}}-1/\eta}  \right) -2 \right) \right].
\end{align}
The original idea of Stoner is to extract the phase diagram of the homogeneous mixture by minimizing equation~(\ref{EtotStoner}) at fixed $n$, resulting in the sample's polarization ${P = |n_1 - n_2|/(n_1 + n_2) = |2 x - 1|}$ as a function of the gas parameter $\eta$ of equation~(\ref{eta}), see figure~\ref{betaM} (right), which identifies the onset of polarization at ${\eta \approx 1.15}$. A very similar value (${\eta_0\approx1.22}$) has been obtained recently with fixed-node QMC calculations utilizing hard- and soft-disk potentials~\cite{Pilati2021arxiv}.

To calculate a phase diagram similar to figure~\ref{betaM} (right) for a trapped mixture, we cannot rely on a universal parameter $\eta$ since the local Fermi energy and the polarization change alongside the local density at fixed fermion number. 
Analogous to studies of mixtures in 3D, we therefore introduce a universal gas parameter
\begin{align}\label{eta0}
\eta_0 =- \frac{1}{\log (k_{\mathrm{F}}^0 a_{\text{2D}})}
\end{align}
for a trapped mixture in 2D, based on the Fermi wavenumber ${k_{\mathrm{F}}^0 = (8 N)^{1/4}/a_{\text{ho}}}$ of the noninteracting system at the center of the trap and the geometric average ${a_{\text{ho}} = \sqrt{\hbar / m \sqrt{\omega_x \omega_y}}}$ of harmonic oscillator lengths.
We compute the total polarization (viz., net magnetization) $\mathcal{P}$ of the trapped mixture as a spatial integral of the local polarization:
\begin{align}\label{polarization}
\mathcal{P}= \frac{\int(\d\VEC r) \left| n_1(\VEC r) - n_2 (\VEC r) \right|}{N}.
\end{align}

\FloatBarrier

\section{\label{Results}Results}

Our primary objective is the reliable prediction of experimentally relevant density distributions for ultracold two-component Fermi gases with contact-type interactions. While DPFT as our principal tool of investigation can be applied to virtually all such settings and beyond, here we focus on harmonically trapped gases in 2D, subjected to (i) the mean-field contact interaction of equation~(\ref{EintPureCon}) and (ii) the renormalized contact interaction of equation~(\ref{EintRenCon}). Its simplicity makes the former a popular approximation, but we find that the (more realistic) renormalized interaction implies quite different density profiles for intermediate interaction strengths. For both choices of the interaction functional and for strong interactions, however, we predict a ferromagnetic state that separates both components into two semi-disks with minimal interface. We also find that by increasing the particle number in the ferromagnetic phase, we diminish the overlap of the two components across the interface, which likely results in decreased dimer formation. In light of our successful benchmarking of DPFT predictions against HF results, we argue that itinerant ferromagnetism on the repulsive branch of the many-body spectrum is a real and robust phenomenon of 2D Fermi gases.

We will begin by demonstrating that the quasi-classical TF approximation, which is supposed to become accurate for large particle numbers, is in fact inapplicable to any particle number for strong interactions. It is therefore necessary to go beyond the TF approximation. Here, we use the systematically quantum-corrected density formula $n_{3'}$, which also proves sufficient as it captures the essential features of Hartree--Fock results for up to ${N=N_1+N_2=110}$ particles (which is close to the practical limit of high-throughput HF calculations): First, we obtain an excellent quantitative agreement of the density profiles $n_{3'}$ and $n_{\mathrm{HF}}$, especially for strong interactions. Second, both methods reveal a transition from the paramagnetic phase at small interaction strengths to intricate partially separated profiles to an almost complete segregation into two semi-disks. Third, we predict similar polarization curves (across particle numbers and with both DPFT and HF) that measure this transition according to equation~(\ref{polarization}). Here, we exhibit a striking disparity to the QMC results for uniform systems, which highlights the crucial influence of the trapping potential.

Our simulations for up to 10000 particles thus present a reliable picture of the phase transitions of the repulsive two-component Fermi gas and inform experimenters about which real-space density profiles to expect across interaction strengths. Of particular importance in this respect are the many qualitatively different configurations of metastable states, which we encounter at intermediate interaction strengths. These metastable density profiles often have energy differences, both relative to each other and to the ground state, of the order of $10^{-6}$--$10^{-3}$ and are therefore likely to be observed in experiments. We detail our findings in the following sections.

\subsection{\label{SectionTF}Lessons from the TF model}
Using the TF-approximated kinetic energy ${E_{\mathrm{kin}}^{\mathrm{TF}}=\frac{c}{2}\int(\d\VEC r)\,\big(n_1^2+n_2^2\big)}$, where ${c=2\pi\hbar^2/(mg)}$, we can analytically solve the two resulting variational equations
\begin{align}
(n_1-n_2)\,(1-\alpha/c)&=(\mu_1-\mu_2)/c\label{TFvareq1}
\intertext{and}
(n_1+n_2)&=(\mu_1+\mu_2-2V_{\mathrm{ext}})/c\label{TFvareq2}
\end{align}
for the mean-field contact interaction of equation~(\ref{EintPureCon}). Setting ${\alpha\not=c}$, we circumvent the fine-tuning problem of equation~(\ref{TFvareq1}), which then yields ${N_1\not=N_2}$ from ${\mu_1\not=\mu_2}$. That is, a balanced mixture ${N_1=N_2}$, implying ${\mu_1=\mu_2}$ and, hence, ${n_1=n_2=n/2}$ with total energy
\begin{align}
E_{\mathrm{TF}}=\int(\d\VEC r)\,\left(\frac{c+\alpha}{4}n^2+V_{\mathrm{ext}}\,n\right)
\end{align}
in 2D does not separate in the TF model---with the proviso that equations~(\ref{TFvareq1}) and (\ref{TFvareq2}) follow from the TF energy functional only if both $n_1$ and $n_2$ are strictly positive. Indeed, at the boundary (${n_i=0}$ and ${n_{j\not=i}=n}$) of the support of $E_{\mathrm{TF}}[n_1,n_2]$, we find
\begin{align}
\tilde{E}_{\mathrm{TF}}=\sum_{i=1,2}\int_{\mathcal D_i}(\d\VEC r)\,\left(\frac{c}{2}n_i^2+V_{\mathrm{ext}}\,n_i\right)=\int(\d\VEC r)\,\left(\frac{c}{2}n^2+V_{\mathrm{ext}}\,n\right)
\end{align}
for arbitrary domains $\mathcal D_1$ and ${\mathcal D_2=\mathbb{R}^2\backslash \mathcal D_1}$ (i.e., ${E_{\mathrm{int}}^{\mathrm{MF}}=0}$) that yield ${N_i=N/2=\int_{\mathcal D_i}(\d\VEC r)\,n_i(\VEC r)}$. Because of ${\tilde{E}_{\mathrm{TF}}<E_{\mathrm{TF}}\Leftrightarrow\alpha>c=2\pi}$ (in harmonic oscillator units), we find a universal (viz., $N$-independent) phase transition at ${\alpha=2\pi}$, beyond which \textit{any} fully polarized phase (characterized by domains $\mathcal D_i$) is energetically favored over the paramagnetic phase that features ${n_1=n_2}$ everywhere; figure~\ref{TFvsn3p} in appendix~\ref{AppendixDPFT} illustrates the analogous situation for the renormalized contact interaction. Therefore, as shown in the following, it is no coincidence that the transition towards the complete split into two semi-disks occurs at ${\alpha\approx2\pi}$ for all particle numbers and is heralded by slow convergence of the intricate phase patterns (with both DPFT and HF), which we encounter for ${\alpha\lesssim2\pi}$. Clearly, the TF approximation cannot adequately describe such a two-component system with repulsive contact interaction, since we could increase the actual kinetic energy $\big($not $E_{\mathrm{kin}}^{\mathrm{TF}}\big)$ of the TF profiles at will by fragmenting the domains $\mathcal D_i$ at fine scales.

\begin{figure}[htb!]
\begin{minipage}{0.195\linewidth}
\begin{center}
${N_{1/2}=10}$
\end{center}
\vspace{-\baselineskip}
\rule{\linewidth}{0.5pt}\\[0.5em]
\includegraphics[height=0.8\linewidth]{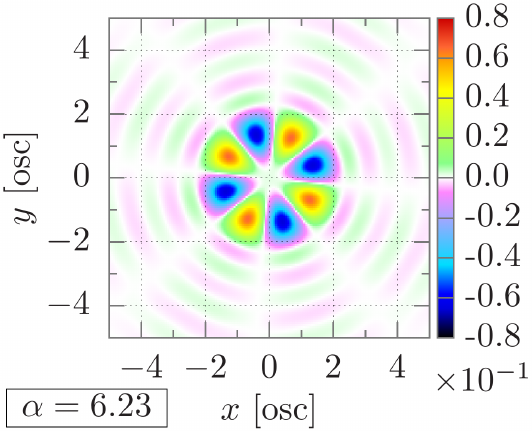}
\includegraphics[height=0.8\linewidth]{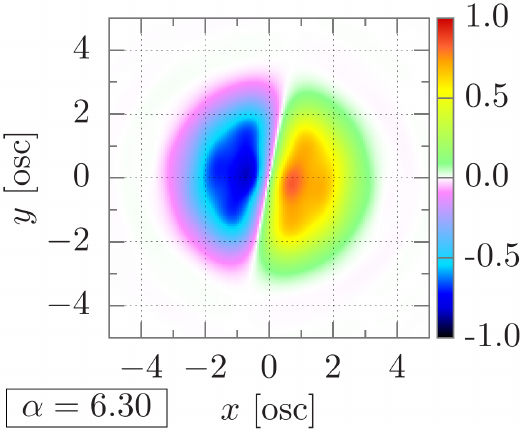}
\end{minipage}
\begin{minipage}{0.195\linewidth}
\begin{center}
${N_{1/2}=15}$
\end{center}
\vspace{-\baselineskip}
\rule{\linewidth}{0.5pt}\\[0.5em]
\includegraphics[height=0.8\linewidth]{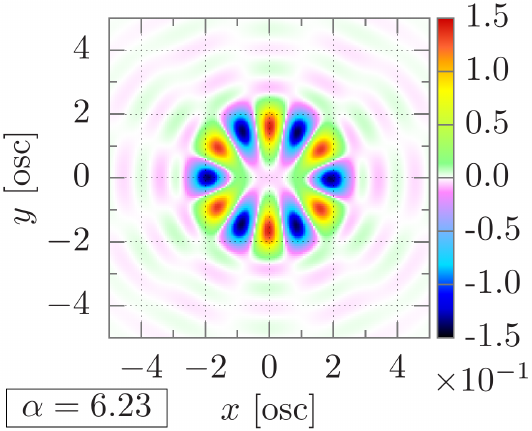}
\includegraphics[height=0.8\linewidth]{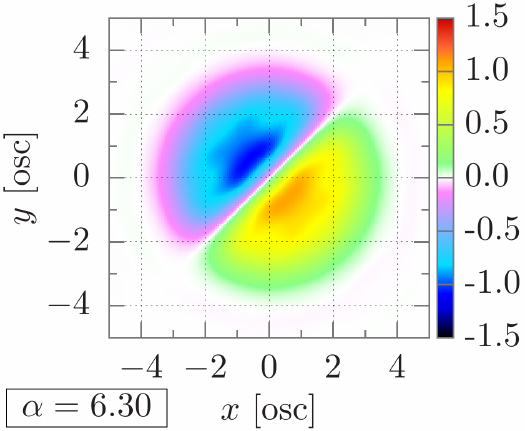}
\end{minipage}
\begin{minipage}{0.195\linewidth}
\begin{center}
${N_{1/2}=55}$
\end{center}
\vspace{-\baselineskip}
\rule{\linewidth}{0.5pt}\\[0.5em]
\includegraphics[height=0.8\linewidth]{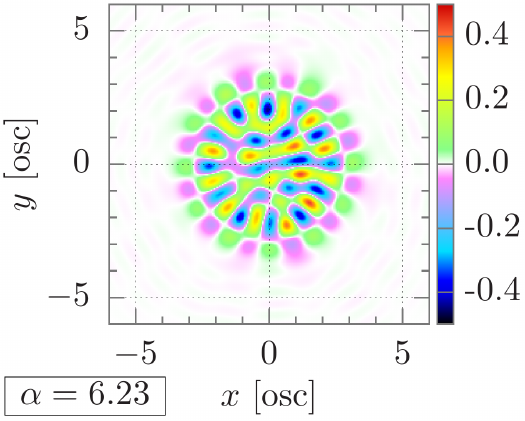}
\includegraphics[height=0.8\linewidth]{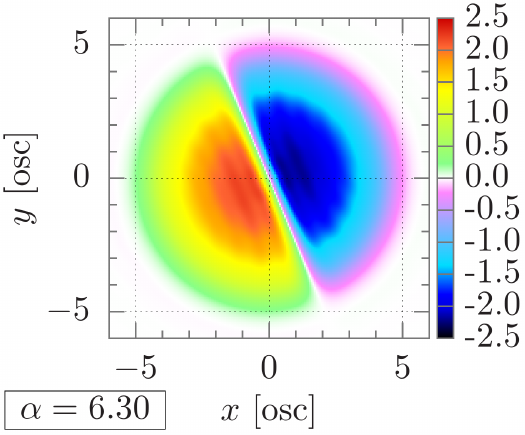}
\end{minipage}
\begin{minipage}{0.195\linewidth}
\begin{center}
${N_{1/2}=500}$
\end{center}
\vspace{-\baselineskip}
\rule{\linewidth}{0.5pt}\\[0.5em]
\includegraphics[height=0.8\linewidth]{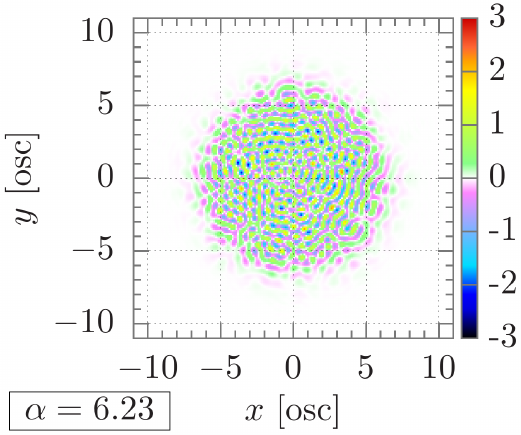}
\includegraphics[height=0.8\linewidth]{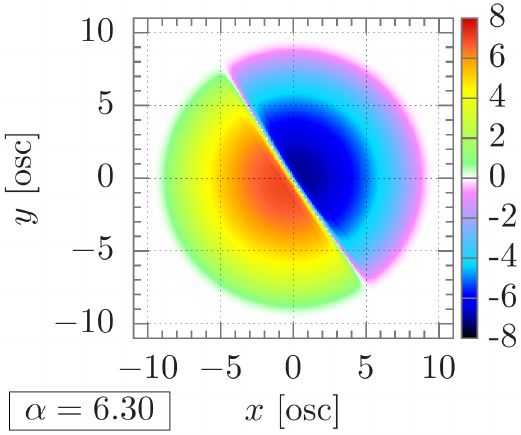}
\end{minipage}
\begin{minipage}{0.195\linewidth}
\begin{center}
${N_{1/2}=5000}$
\end{center}
\vspace{-\baselineskip}
\rule{\linewidth}{0.5pt}\\[0.5em]
\includegraphics[height=0.8\linewidth]{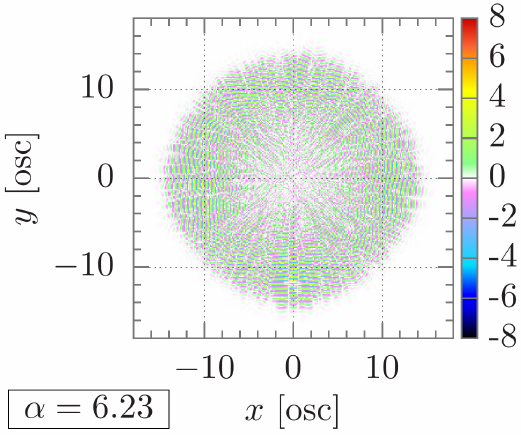}
\includegraphics[height=0.8\linewidth]{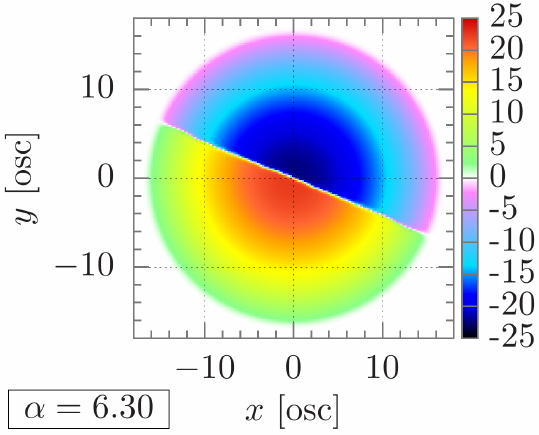}
\end{minipage}
\caption{\label{PureContactInteraction}The ground-state ($n_{3'}$-)approximated local polarizations ${n_1-n_2}$ for the case of the mean-field contact-interaction functional of equation~(\ref{EintPureCon}). Up to ${\alpha\lesssim6.2\,(<2\pi)}$ both density profiles are identical (paramagnetic phase). Strong repulsive interactions, that is, ${\alpha\gtrsim6.3\,(>2\pi)}$, separate both species into two semi-disks (ferromagnetic phase), encircled by a ring of almost identical profiles, akin to the outer region of the cross-sections shown for the renormalized contact interaction in figure~\ref{CompareCrossSection} below.}
\end{figure}

Since a nonlocal treatment of the kinetic energy is necessary, we will deploy the semiclassical DPFT framework with the density formulae $n_{3'}$ of equations~(\ref{n3p}) and (\ref{n3pF}). Indeed, $n_{3'}$ delivers a particle-number-independent transition from the paramagnetic phase at ${\alpha\lesssim6.2}$ across ${\alpha=2\pi}$ towards a ferromagnetic phase for ${\alpha\gtrsim6.3}$, see figure~\ref{PureContactInteraction}. The separation into two domains comes with minimal interface, as one would intuitively expect for strong enough repulsion. In between the para- and ferromagnetic states, we observe a zoo of particle-number-dependent phases; see reference~\cite{Hue2020a} for a comprehensive set of plots beyond those shown in figure~\ref{PureContactInteraction}. As the particle number increases, the density profiles at ${\alpha<2\pi}$ become increasingly fragmented into partially spin-polarized domains, with many different domain configurations that are close in energy. Consequently, the convergence of the self-consistent DPFT loop in equation~(\ref{SCloop}) requires a large number of iterations at high spatial resolution and high numerical accuracy. In fact, for ${N_s\ge55}$ at ${\alpha=6.23}$ we find the domains in a state of perpetual transformation: The three according plots in figure~\ref{PureContactInteraction} are snapshots after ${\sim10^5}$ iterations.

These phase transitions of the contact-interacting two-component Fermi gas in 2D are clearly different from (and more diverse than) their 3D counterparts, which merely evolve from a symmetric phase to a splitting into two semi-spheres through an intermediate isotropic separation \cite{Trappe2016}. Also the transition window (${6.2\lesssim\alpha\lesssim6.3}$) contrasts with the 3D situation, where the phase transition sets in (and is completed) at smaller $\alpha$ for larger $N_{1/2}$. Furthermore, when constraining the densities to isotropic profiles, we find no separation at all for $\alpha<2\pi$, with energies slightly above those of the anisotropic ground-state separations shown in figure~\ref{PureContactInteraction}.

In summary, DPFT yields a transition into the ferromagnetic phase at ${\alpha=2\pi}$ for both $n_{\mathrm{TF}}$ and $n_{3'}$. Unlike $n_{\mathrm{TF}}$, however, which does not predict any segregations for ${\alpha<2\pi}$, $n_{3'}$ segregates the two fermion components into intricate patterns in a small window below ${\alpha=2\pi}$. We demonstrate in section~\ref{RenCon} below that the (more realistic) renormalized rather than the mean-field contact interaction should be deployed for reliable simulations since the respective density profiles of this intermediate phase differ markedly.

\subsection{\label{DPFTvsHF}Benchmarking DPFT against Hartree-Fock}

In order to gain confidence in the DPFT predictions for the mesoscopic particle numbers realized in experiments on contact-interacting ultracold Fermi gases, we now benchmark our $n_{3'}$-based density profiles against HF calculations for particle numbers up to ${N_1+N_2=110}$.

\begin{figure}[htb!]
\includegraphics[width=0.28\linewidth]{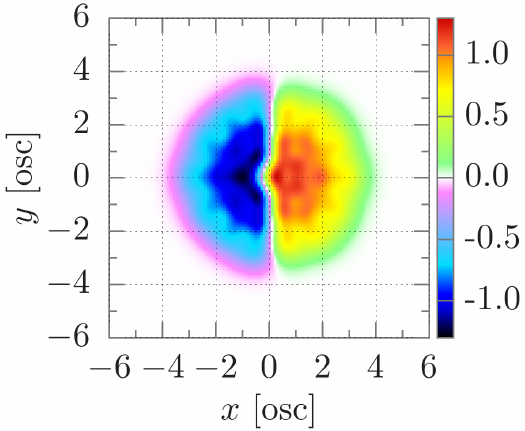}
\hfill
\includegraphics[width=0.28\linewidth]{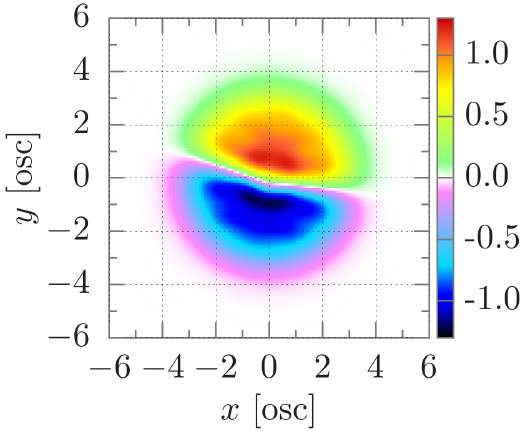}
\hfill
\includegraphics[width=0.33\linewidth]{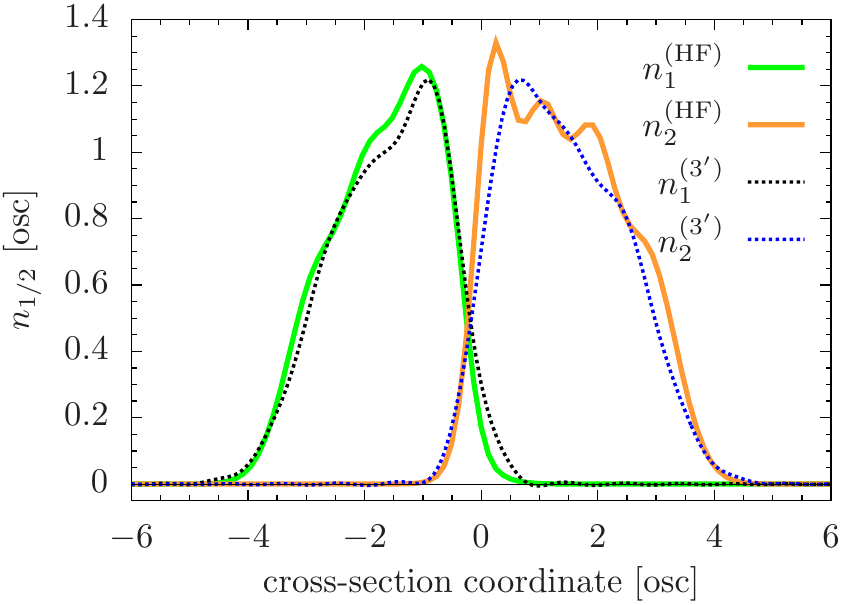}\\[\baselineskip]
\includegraphics[width=0.28\linewidth]{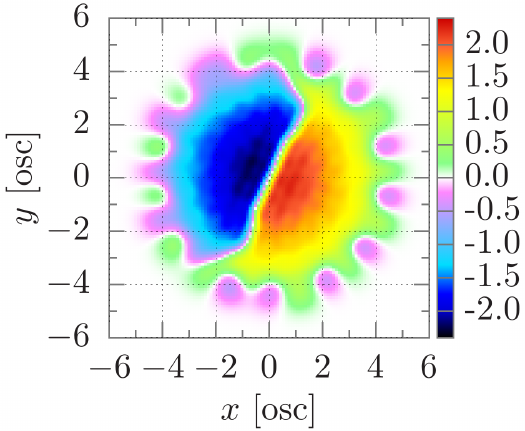}
\hfill
\includegraphics[width=0.28\linewidth]{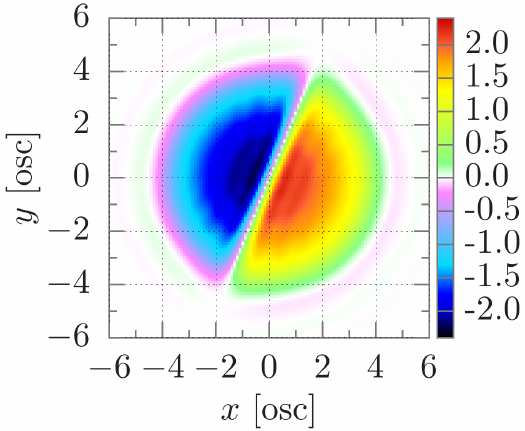}
\hfill
\includegraphics[width=0.33\linewidth]{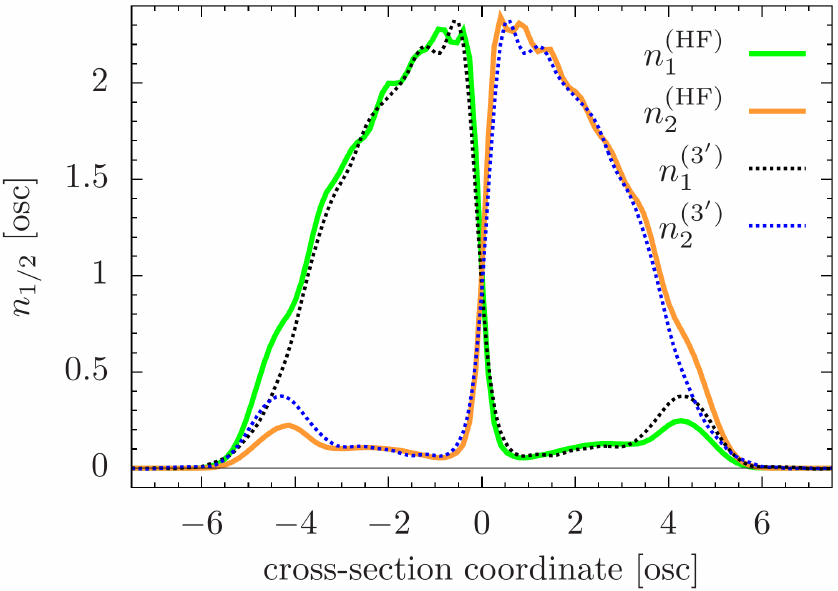}\\[\baselineskip]
\includegraphics[width=0.28\linewidth]{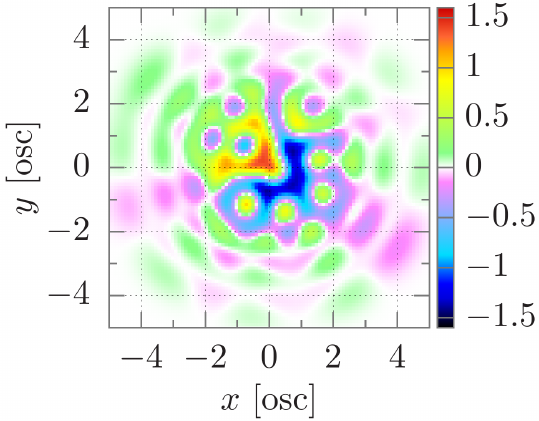}
\hfill
\includegraphics[width=0.28\linewidth]{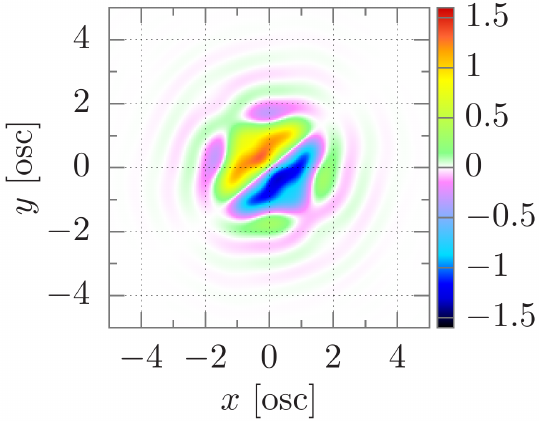}
\hfill
\includegraphics[width=0.33\linewidth]{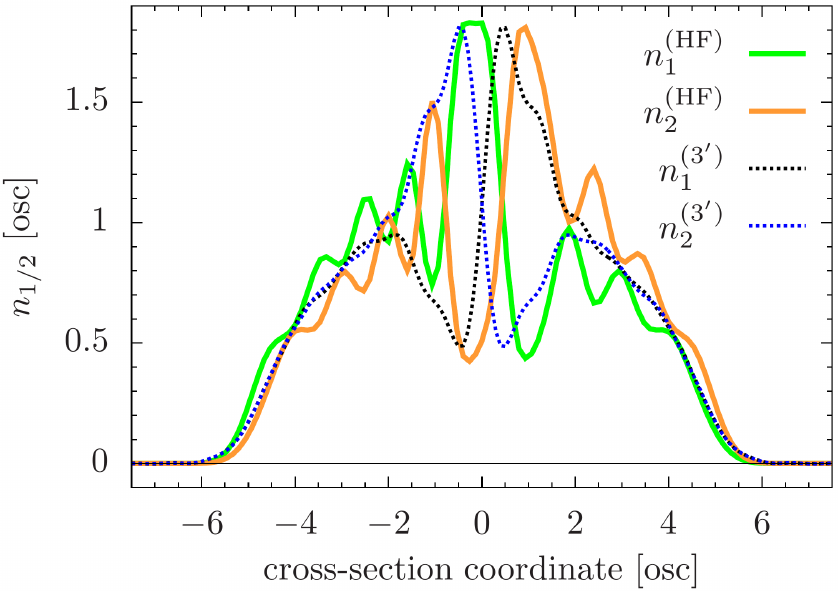}\\[\baselineskip]
\caption{\label{CompareCrossSection}We find excellent agreement between the density profiles $n_{1/2}$ obtained from DPFT with $n_{3'}$ and those obtained from HF. We depict (i) the local polarizations ${n_1-n_2}$ for HF (left column) and DPFT (middle column) and (ii) the cross sections of $n_{1/2}$ along the axis perpendicular to the interface (right column). Top row: $N_{1/2}=15$, mean-field contact interaction strength ${\alpha=8}$. Middle (bottom) row: $N_{1/2}=55$, renormalized contact interaction strength ${\eta_0=9}$ (${\eta_0=2.2}$).} 
\end{figure}

\FloatBarrier

Figure~\ref{CompareCrossSection} demonstrates that $n_{3'}$ captures all essential features of the HF densities, both qualitatively and quantitatively. This includes overall structure, cloud sizes, evanescent tails, as well as extent and gradient of the interface between the two fermionic species. The latter property is of particular importance for reliable estimates of dimer formation rates~\cite{Valtolina2016}. Although $n_{3'}$ even predicts nontrivial HF features like the marginally asymmetric separation depicted in the top row of figure~\ref{CompareCrossSection}, the similar feature of $n_{\mathrm{HF}}$ in the middle row is not captured. There, however, $n_{3'}$ accurately exhibits the partial polarization magnitudes of $n_{\mathrm{HF}}$ as well as the revival of the minority component in the outer cloud ring. That is, $n_{3'}$ presents a reliable density expression at least for strong interactions that clearly yield a ferromagnetic state in both the HF and DPFT framework. This close match between both methods is less pronounced for intermediate interaction strengths, where the complex structures in the outer cloud regions are somewhat disparate and only HF predicts an asymmetric separation (bottom row of figure~\ref{CompareCrossSection}). As we will discuss in the next section, the regime of intermediate interaction strengths is the richest in terms of possible metastable configurations that are energetically comparable and poses, therefore, numerically challenging for both HF and DPFT.

\subsection{\label{RenCon}Phase transitions and total polarizations of Fermi gases with renormalized contact interaction}

In section~(\ref{DPFTvsHF}), we established the high quality of the quantum-corrected DPFT formula $n_{3'}$, see equations~(\ref{n3p}) and (\ref{n3pF}), by benchmarking against HF results for mean-field and renormalized contact interactions for moderate particle numbers. We now present the main results of this work for balanced mixtures with ${N=N_1+N_2}$ up to $10000$ as required for describing realistic experimental setups~\cite{Giorgini2008}.

\begin{figure}[htb!]
\begin{minipage}{0.195\linewidth}
\begin{center}
${N_{1/2}=10}$
\end{center}
\vspace{-\baselineskip}
\rule{\linewidth}{0.5pt}\\[0.5em]
\includegraphics[height=0.8\linewidth]{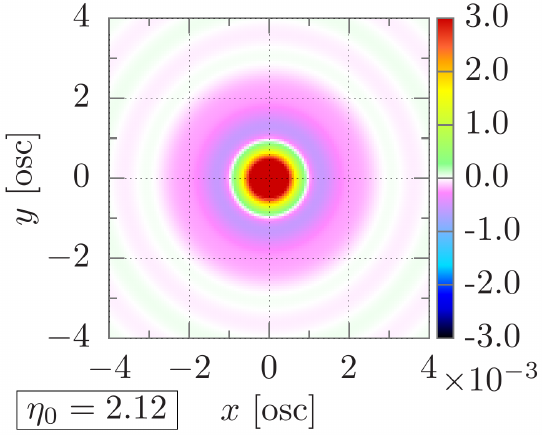}
\includegraphics[height=0.8\linewidth]{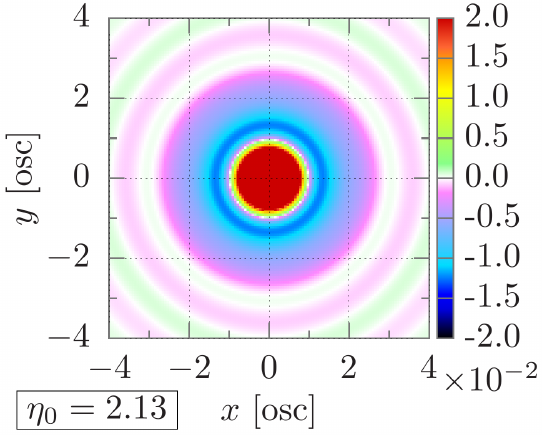}
\includegraphics[height=0.8\linewidth]{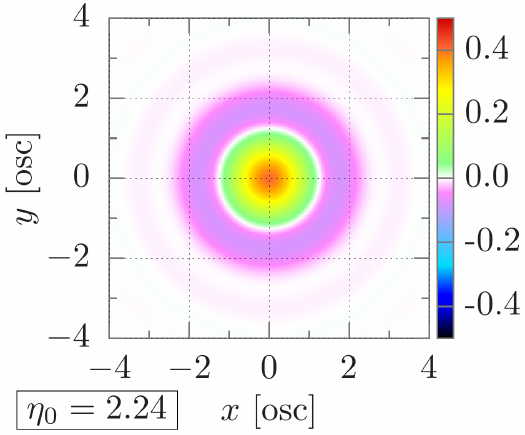}
\includegraphics[height=0.8\linewidth]{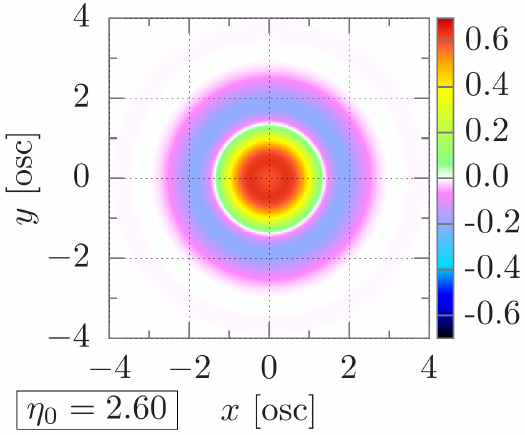}
\includegraphics[height=0.8\linewidth]{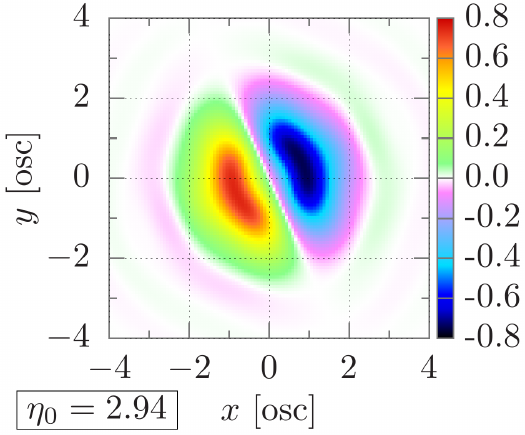}
\includegraphics[height=0.8\linewidth]{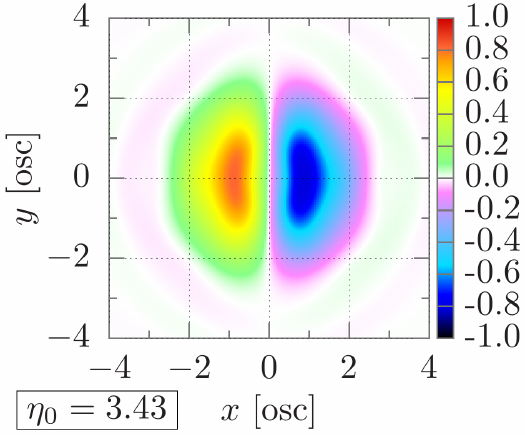}
\includegraphics[height=0.8\linewidth]{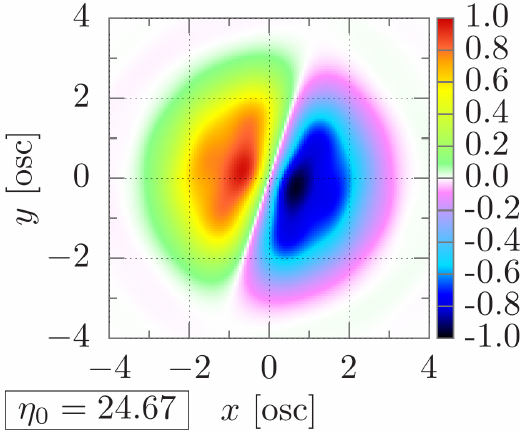}
\end{minipage}
\begin{minipage}{0.195\linewidth}
\begin{center}
${N_{1/2}=15}$
\end{center}
\vspace{-\baselineskip}
\rule{\linewidth}{0.5pt}\\[0.5em]
\includegraphics[height=0.8\linewidth]{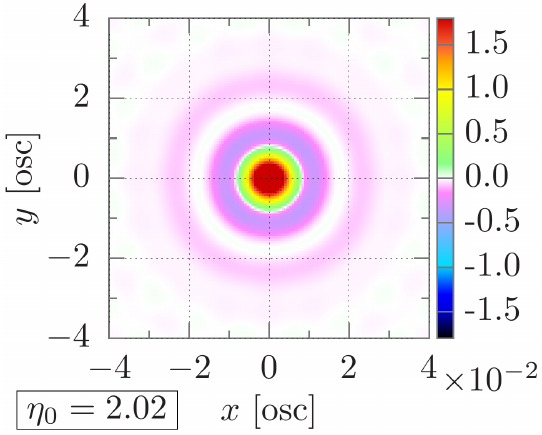}
\includegraphics[height=0.8\linewidth]{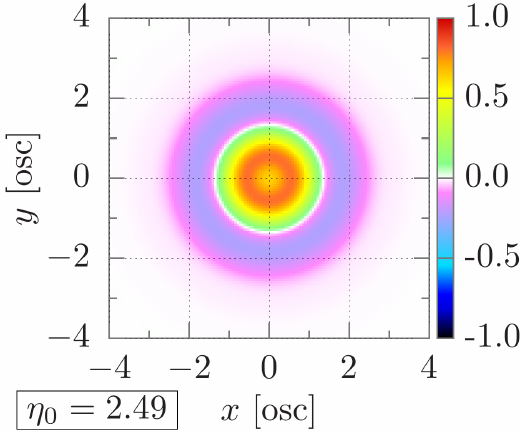}
\includegraphics[height=0.8\linewidth]{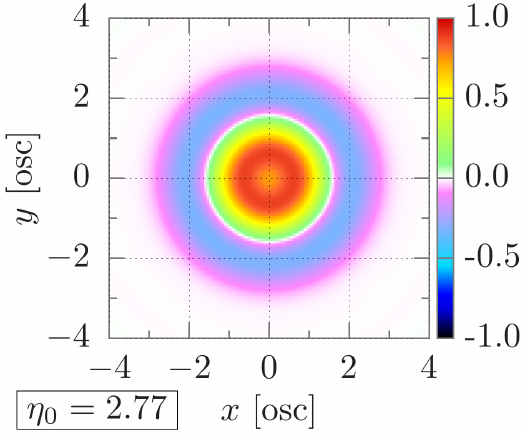}
\includegraphics[height=0.8\linewidth]{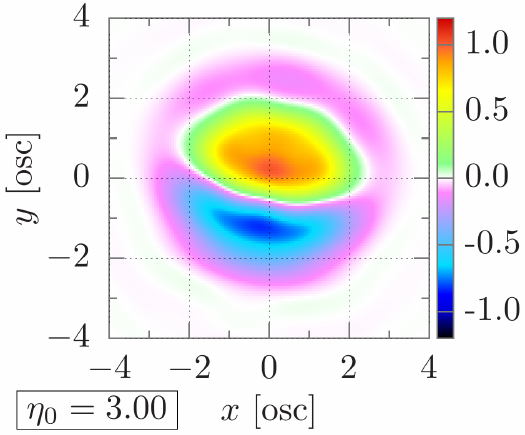}
\includegraphics[height=0.8\linewidth]{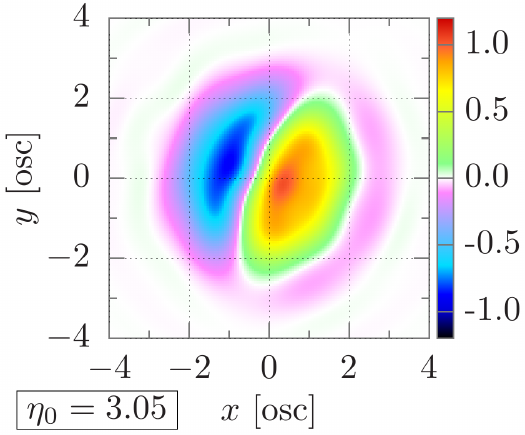}
\includegraphics[height=0.8\linewidth]{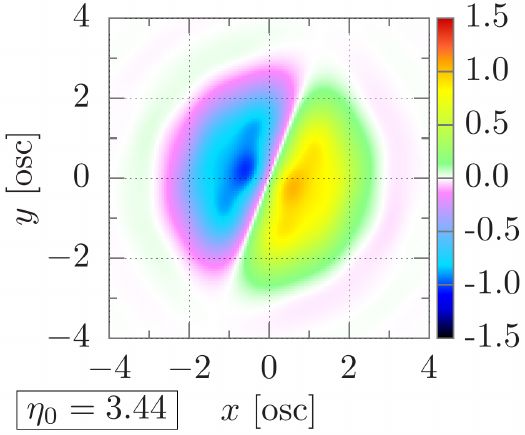}
\includegraphics[height=0.8\linewidth]{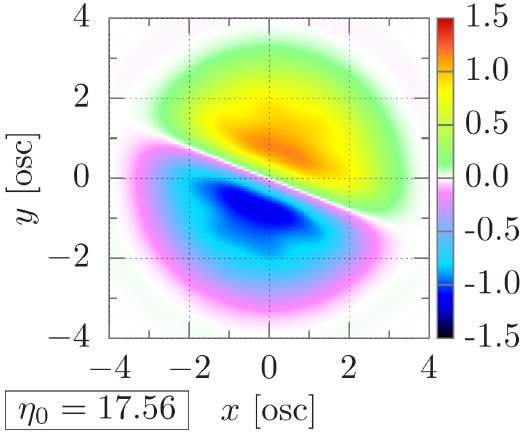}
\end{minipage}
\begin{minipage}{0.195\linewidth}
\begin{center}
${N_{1/2}=55}$
\end{center}
\vspace{-\baselineskip}
\rule{\linewidth}{0.5pt}\\[0.5em]
\includegraphics[height=0.8\linewidth]{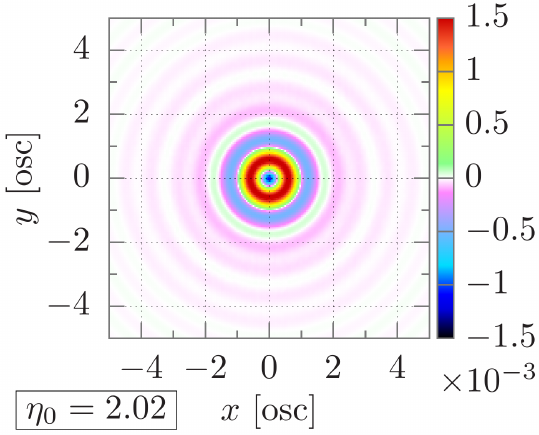}
\includegraphics[height=0.8\linewidth]{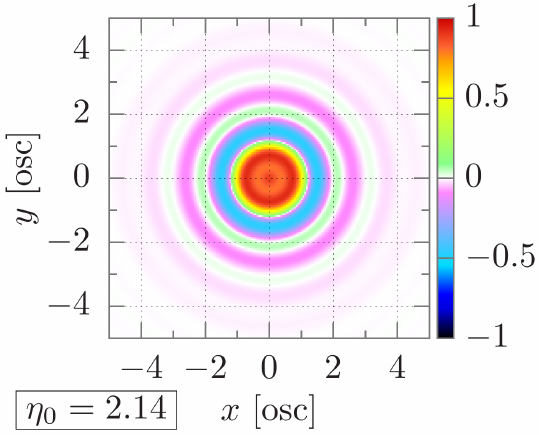}
\includegraphics[height=0.8\linewidth]{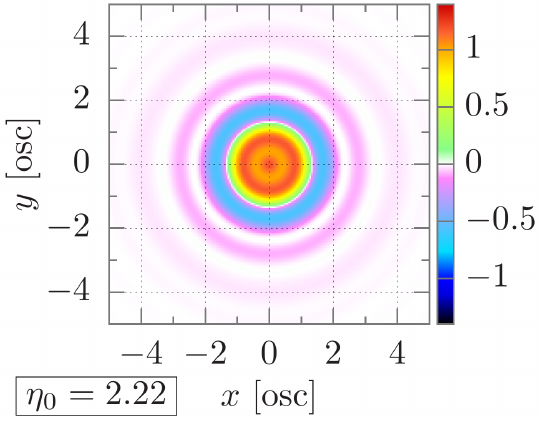}
\includegraphics[height=0.8\linewidth]{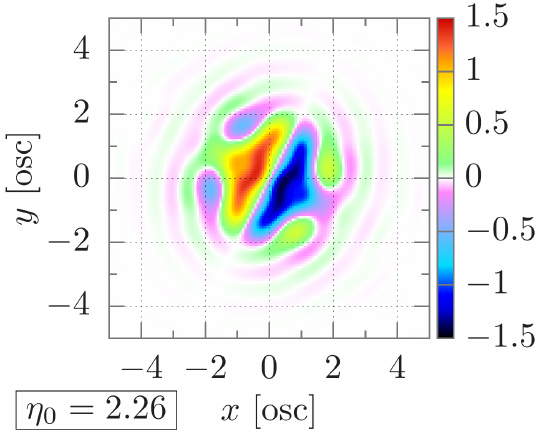}
\includegraphics[height=0.8\linewidth]{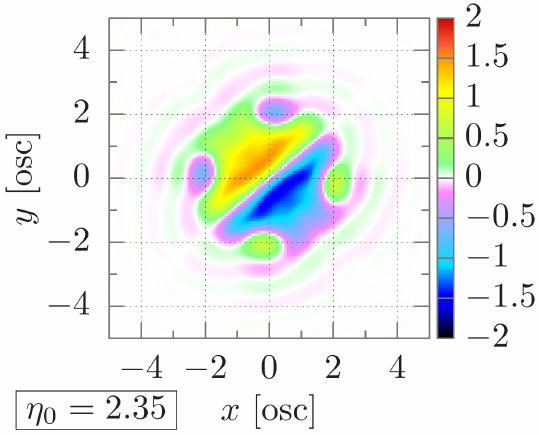}
\includegraphics[height=0.8\linewidth]{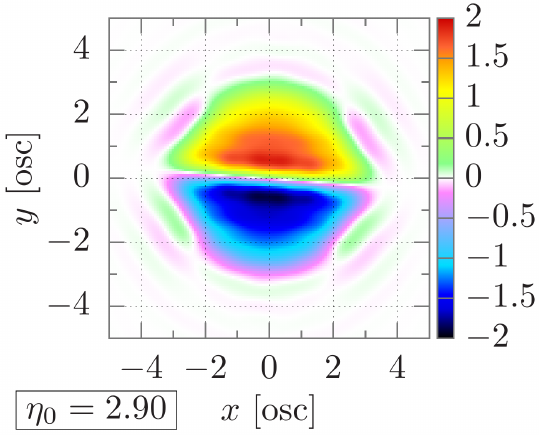}
\includegraphics[height=0.8\linewidth]{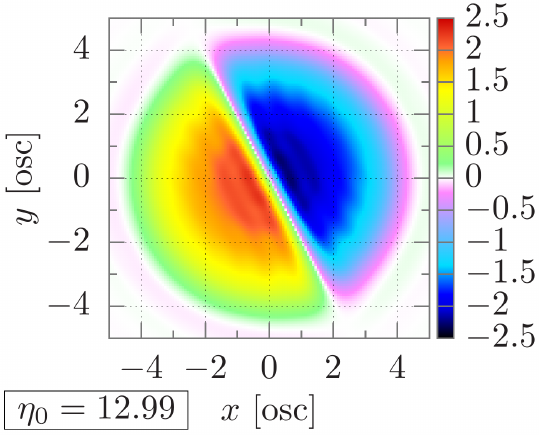}
\end{minipage}
\begin{minipage}{0.195\linewidth}
\begin{center}
${N_{1/2}=500}$
\end{center}
\vspace{-\baselineskip}
\rule{\linewidth}{0.5pt}\\[0.5em]
\includegraphics[height=0.8\linewidth]{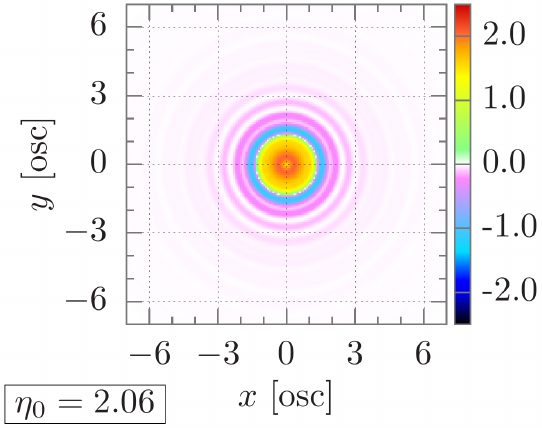}
\includegraphics[height=0.8\linewidth]{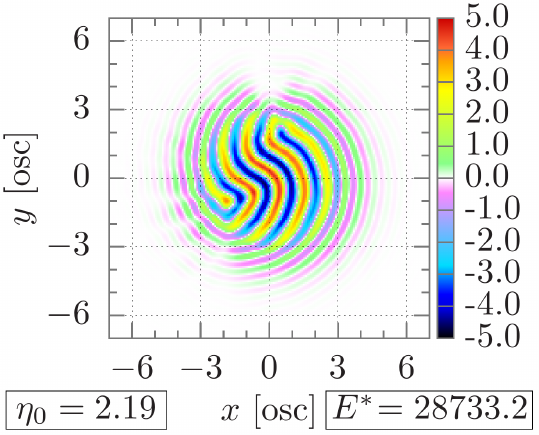}
\includegraphics[height=0.8\linewidth]{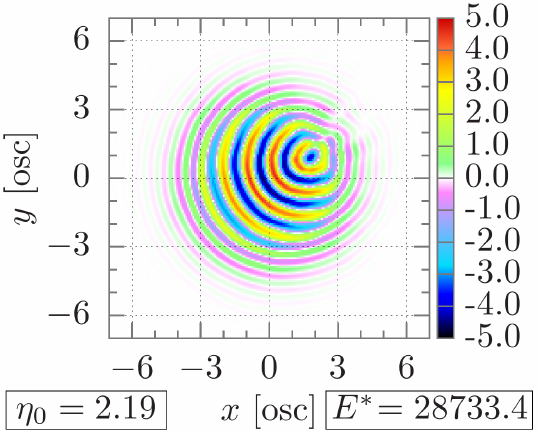}
\includegraphics[height=0.8\linewidth]{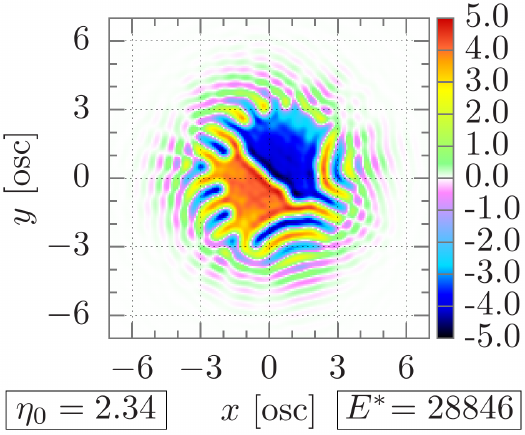}
\includegraphics[height=0.8\linewidth]{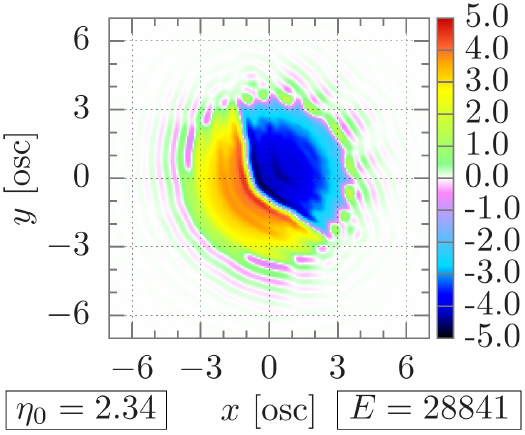}
\includegraphics[height=0.8\linewidth]{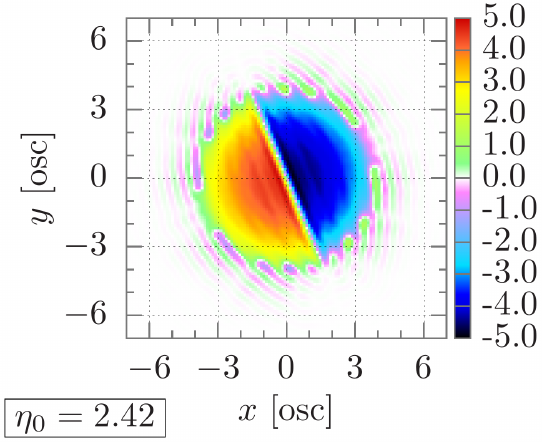}
\includegraphics[height=0.8\linewidth]{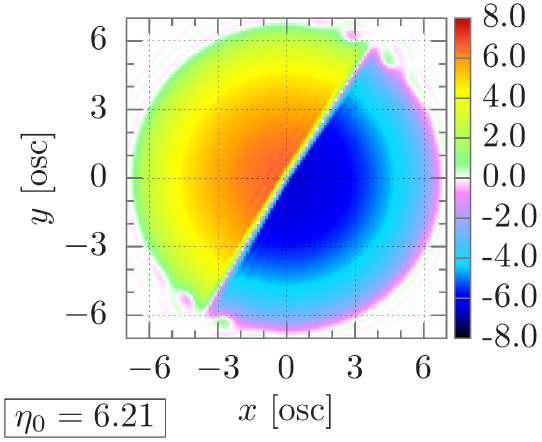}
\end{minipage}
\begin{minipage}{0.195\linewidth}
\begin{center}
${N_{1/2}=5000}$
\end{center}
\vspace{-\baselineskip}
\rule{\linewidth}{0.5pt}\\[0.5em]
\includegraphics[height=0.8\linewidth]{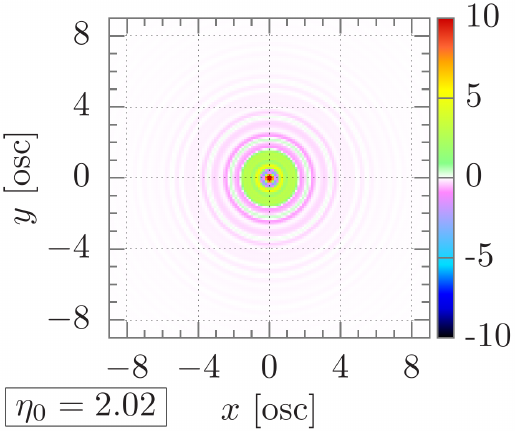}
\includegraphics[height=0.8\linewidth]{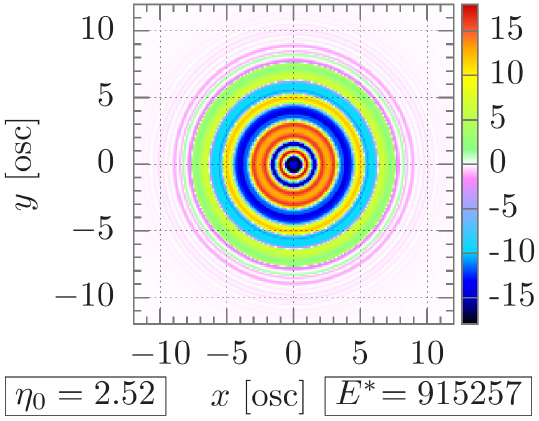}
\includegraphics[height=0.8\linewidth]{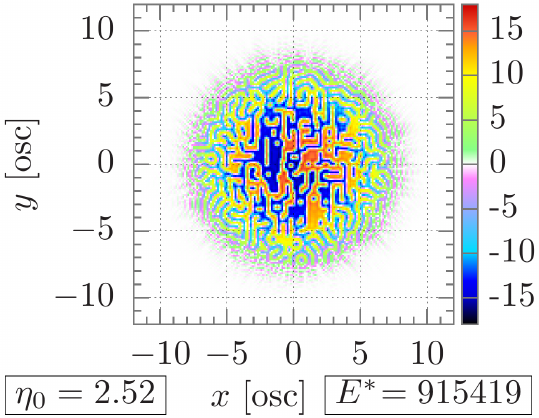}
\includegraphics[height=0.8\linewidth]{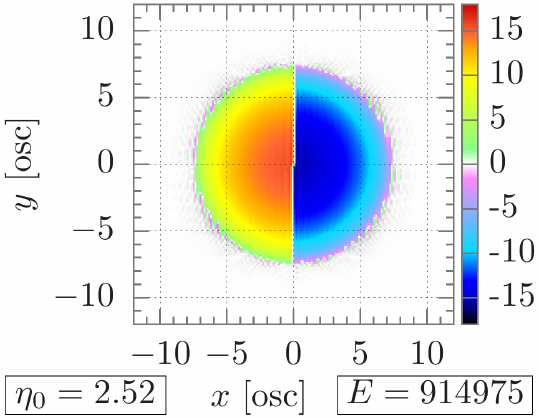}
\includegraphics[height=0.8\linewidth]{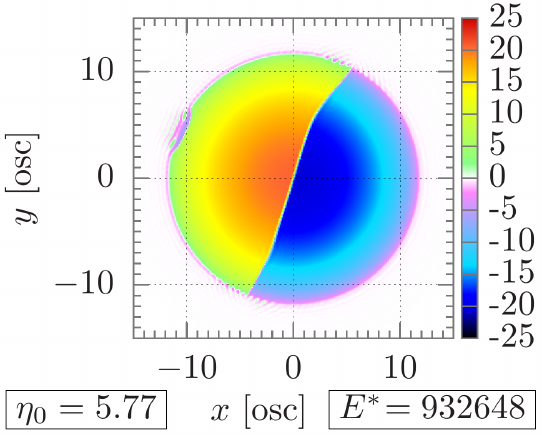}
\includegraphics[height=0.8\linewidth]{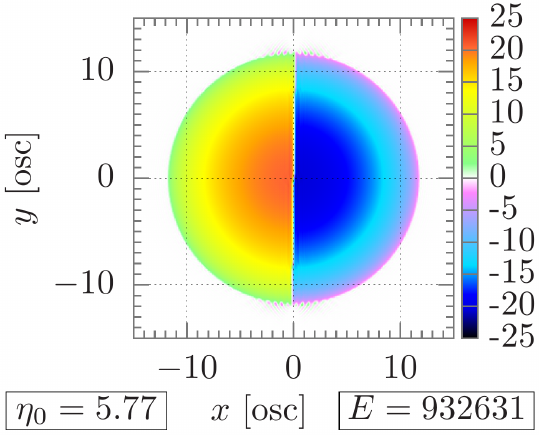}
\includegraphics[height=0.8\linewidth]{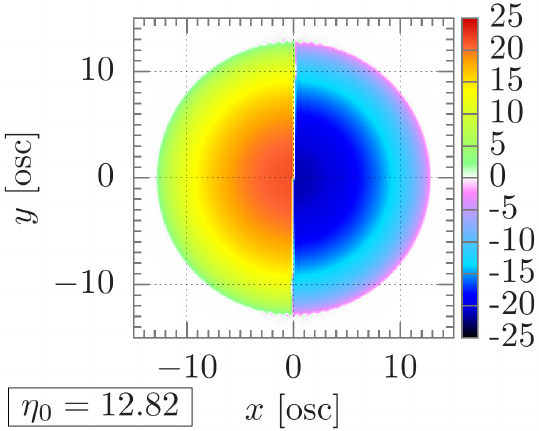}
\end{minipage}
\caption{\label{RenConDensityDiffn3p}Local polarizations ${n_1-n_2}$ of two-component Fermi gases subjected to the renormalized contact interaction of equation~(\ref{EintRenCon}). Each column for fixed particle numbers $N_{1/2}$ represents a sequence of density profiles for increasing interaction strength (from top to bottom) in terms of the universal gas parameter $\eta_0$. The $n_{3'}$-densities calculated within the DPFT framework reveal a ferromagnetic configuration with minimal interface between both Fermi components if ${\eta_0>\eta_0^{\mathrm c}}$, where the critical interaction strength $\eta_0^{\mathrm c}$ lies approximately between $2.0$ and $3.1$, with a tendency of smaller $\eta_0^{\mathrm c}$ for larger particle number; for example, $2.0\lesssim\eta_0^{\mathrm c}\lesssim2.1$ for $N_s=5000$. We state the ground-state energies $E$ when comparing with metastable/transient configurations ($E^*$) at the same $\eta_0$.}
\vspace*{-20pt}
\end{figure}

Figure~\ref{RenConDensityDiffn3p} shows two universal features of the local polarizations $n_1-n_2$ in $n_{3'}$-approximation across various balanced mixtures (${N_1=N_2}$) and interaction strengths (viz., values of the universal gas parameter $\eta_0$). First, $\eta_0\lesssim2$ implies a paramagnetic phase, where both fermion components have the same density profile. Second, when exceeding a critical interaction strength $\eta_0^{\mathrm c}$, the two components transit into a bipartite splitting towards an almost complete separation into two semi-disks, whose interface becomes more pronounced with increasing particle number. Both features are in line with our results on the mean-field-interacting gas, see figure~\ref{PureContactInteraction} and reference~\cite{Hue2020a}. The depicted sequences (columns in figure~\ref{RenConDensityDiffn3p}) of plots for fixed $N_{1/2}$ are snapshots of the para- to ferromagnetic phase transition.

The noise imposed on the densities at the start of the self-consistent loop of equation~(\ref{SCloop}) breaks the spherical symmetry and lets the two fermion clouds equilibrate with a random orientation that differs from run to run, if $\eta_0$ is large enough to induce anisotropic ground-state separations. The smooth density profiles presented here cannot be obtained using the TF approximation (see also figure~\ref{TFvsn3p}) and are distinctly different from the profiles for the mean-field contact interaction, see figure~\ref{PureContactInteraction}. Figure~\ref{RenConDensityDiffn3p} shows a rich zoo of particle-number-dependent phases at intermediate interaction strengths, from isotropic as well as symmetry-broken partial separations to intricate domain wall structures for larger particle numbers. That is, realistic descriptions of contact-interacting two-component Fermi gases in 2D require interaction density functionals that supersede the mean-field kernel $\alpha\,n_1\,n_2$ of equation~(\ref{EintPureCon}).

As we increase the particle number beyond $\sim100$, we find more and more qualitatively disparate phases with minute energy differences and increasingly intricate separation patterns in the window ${2\lesssim\eta_0\lesssim3}$. For ${N\gtrsim1000}$, the unambiguous identification of the ground state densities is thus both more cumbersome and less relevant since the ground state will less likely be encountered in experiments. For example, the two metastable configurations for $N_{1/2}=500$ at ${\eta_0=2.19}$ have energies of ${E=28733.2}$ and ${E=28733.4}$, close to the ${E=28727.7}$ for the ground state configuration, which is an isotropic separation that emerges smoothly from the configuration at ${\eta_0=2.06}$. These two pictures are snapshots of two separate runs of the self-consistent DPFT loop taken after 30000 and 100000 iterations, respectively. Evidently, the intermediate interaction regime easily promotes transformations between energetically comparable, though qualitatively disparate, density profiles during equilibration. We have made similar observations with the HF calculations for smaller particle numbers and expect that such transient states can also be seen in the laboratory. When employing limited numerical accuracy, one cannot determine whether configurations that are energetically above the ground-state energy are (i) transient or (ii) present a metastable state of a local minimum, even if the self-consistent DPFT loop has converged. In section~\ref{MetastableDensities}, we will illuminate how to identify metastability within our DPFT approach and showcase further examples that complement the global picture of the phase transition displayed in figure~\ref{RenConDensityDiffn3p}.

Finally, we reach the regime of mesoscopic particle numbers (${N_{1/2}=5000}$, shown in the last column of figure~\ref{RenConDensityDiffn3p}) that are commonly studied in ultracold-gas experiments with (quasi-)2D geometries. The para- to ferromagnetic transition parallels what we find for smaller particle numbers, but we expose important aspects that are difficult to extrapolate from the results on small ${N\lesssim100}$. Most importantly, the sharp and seemingly random interfaces that emerge during the self-consistent DPFT equilibration are reminiscent of the `fragmented' TF regime of degenerate domain structures. Indeed, even a substantial restructuring of the domains has only marginal impact on the energy. It is thus not surprising that the convergence towards the ground state profiles demands considerable computational effort. We thus present, as for the case of $N_{1/2}=500$, metastable configurations whose energies are slightly above those of the isotropic ground state for ${\eta_0\lesssim\eta_0^{\mathrm c}}$. Even in the ferromagnetic regime, the DPFT equilibration can get stuck despite various measures that aid the convergence (see section~\ref{MetastableDensities} for details). Therefore, and for the sole purpose of obtaining a clean ferromagnetic separation, we converge with an inter-specific Coulomb interaction of strength $\gamma$ superimposed onto the renormalized contact interaction, followed by an equilibration with gradual reduction of $\gamma$, see appendix~\ref{AppendixDPFT} for details. The result is a symmetrically split configuration with slightly lower energy than configurations whose domain walls are marginally shifted relative to the ground-state configuration. As an example, we present such a situation in figure~\ref{RenConDensityDiffn3p} for ${\eta_0=5.77}$. We verified for small particle numbers that the intricate separation patterns at intermediate interaction strengths are recovered when this artificial interaction is gradually switched off.

\FloatBarrier

We recognize \textit{in retrospect} that the TF model proves qualitatively correct regarding key properties for large particle numbers: The TF densities of mesoscopic mixtures subjected to the mean-field contact interaction share three characteristic features with the quantum-corrected densities for both the mean-field and the renormalized contact interaction. First, the para- to ferromagnetic transition becomes relatively sharp as the particle number increases, both in terms of the 2D scattering length and in terms of $\eta_0$. Second, the ferromagnetic TF solutions, which are degenerate for any domain fragmentation, are structurally similar to the metastable $n_{3'}$ configurations that we encounter for ${N_s=5000}$---figure~\ref{RenConDensityDiffn3p} shows two such examples at ${\eta_0=2.52}$ and ${\eta_0=5.77}$, respectively. Third, the density gradients at the domain interfaces of the quantum-corrected mixtures increase with particle number. Judging from our simulations for up to 10000 particles, we expect this trend to continue toward the quasi-classical limit embodied by the TF model, which features vertical interfaces. 

\begin{figure}[ht!]
\includegraphics[width=0.65\linewidth]{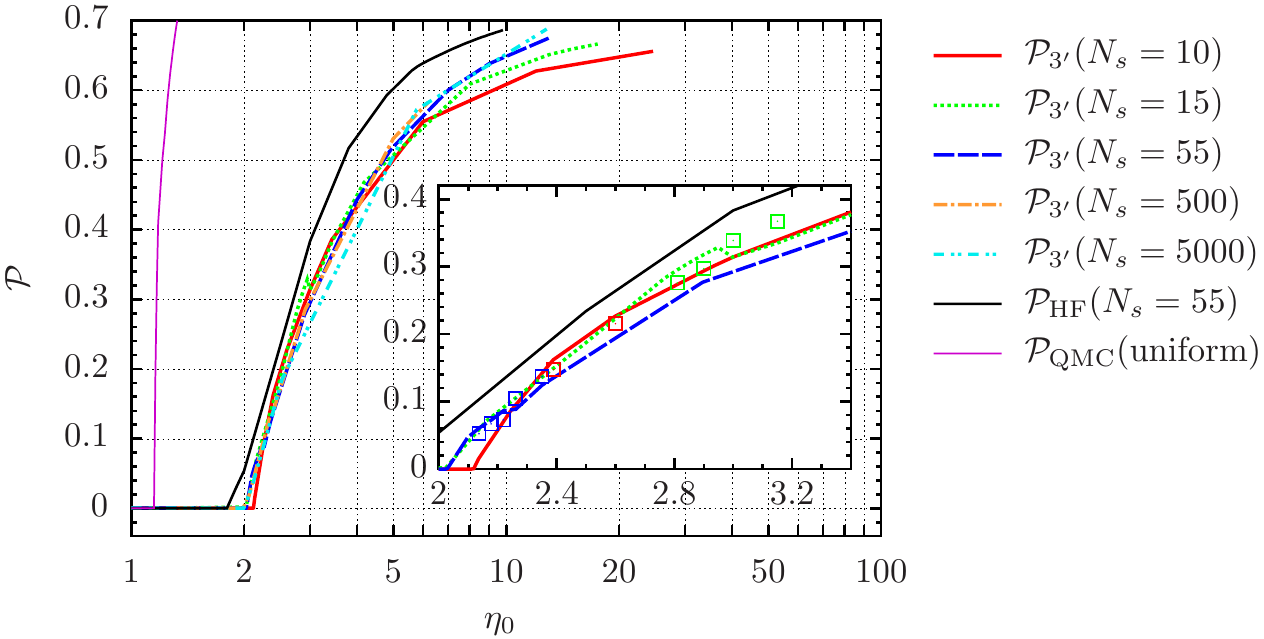}
\caption{\label{Magnetizationn3pFFT}The polarization ($\mathcal P$-)curves for different particle numbers collapse onto a unique function of the universal gas parameter $\eta_0$. In particular, we find a universal onset of phase separation across particle numbers. Accepting the disparities between the $\mathcal P$-curves of different $N_s$ as error estimates of our DPFT approach, we find the polarizations of the metastable states (marked by squares with color code of the respective $\mathcal P$-curves; displayed in figure~\ref{MetastableProfiles}) to approximately equal those of the ground states, but they are located near kinks of the $\mathcal P$-curves that indicate the transition into the ferromagnetic state at $\eta_0^{\mathrm c}$. The inset magnifies the according region of the phase diagram.}
\end{figure}

The universal character of the phase transitions for the harmonically confined fermion mixture illustrated in figure~\ref{RenConDensityDiffn3p} is summarized with figure~\ref{Magnetizationn3pFFT}, which shows the polarization $\mathcal P$ of equation~(\ref{polarization}) as a function of the universal gas parameter $\eta_0$ of equation~(\ref{eta0}). Both the onset of phase separation and a dampened increase towards full polarization are universal across particle numbers and follow qualitatively the homogeneous case shown in figure~\ref{betaM}. The crossing of metastable- and ground-state polarizations shown in the inset of figure~\ref{Magnetizationn3pFFT} parallels the crossing of the respective energies near $\eta_0^{\mathrm c}$ (see also figure~\ref{MetastableProfiles}), where the mixtures transit into the ferromagnetic phase. For example, for ${N_{1/2}=15}$ around ${\eta_0=3}$, the polarization curves of the metastable states, which exhibit anisotropic (isotropic) separations for ${\eta_0\lesssim3}$ (${\eta_0\gtrsim3}$), cross the polarization curves of the ground states, which exhibit isotropic (anisotropic) separations for ${\eta_0\lesssim3}$ (${\eta_0\gtrsim3}$).

\FloatBarrier

\subsection{\label{MetastableDensities}Metastable density profiles}
\begin{figure}[htb!]
\begin{minipage}{\linewidth}
\begin{minipage}{0.3\linewidth}
\includegraphics[width=\linewidth]{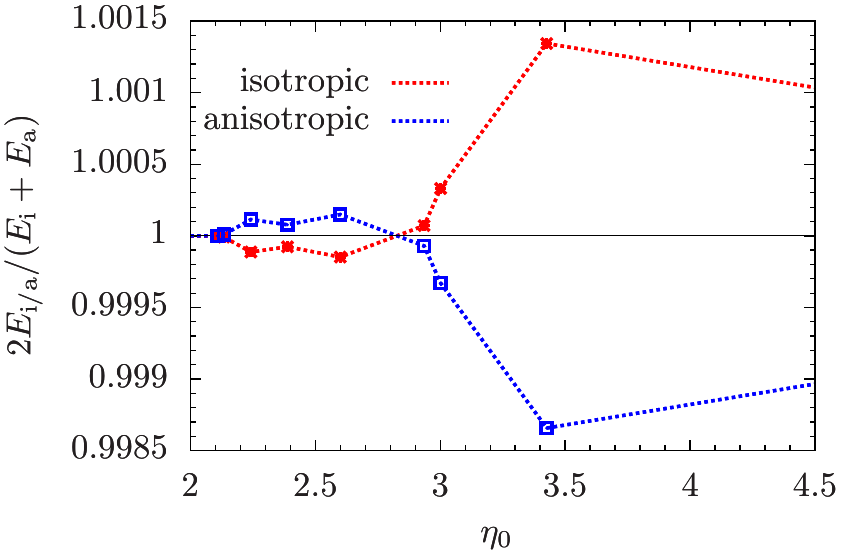}
\end{minipage}
\hspace{0.01\linewidth}
\begin{minipage}{0.131\linewidth}
\begin{center}
${N_{1/2}=10}$
\end{center}
\vspace{-1.5\baselineskip}
\rule{\linewidth}{0.5pt}\\[0.5em]
\includegraphics[width=\linewidth]{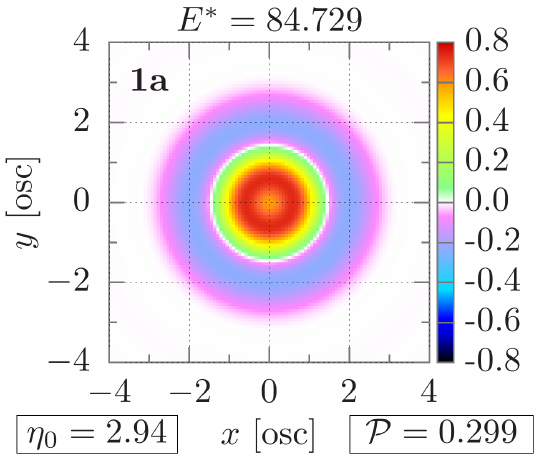}
\phantom{x}\vspace{-0.5em}
\includegraphics[width=\linewidth]{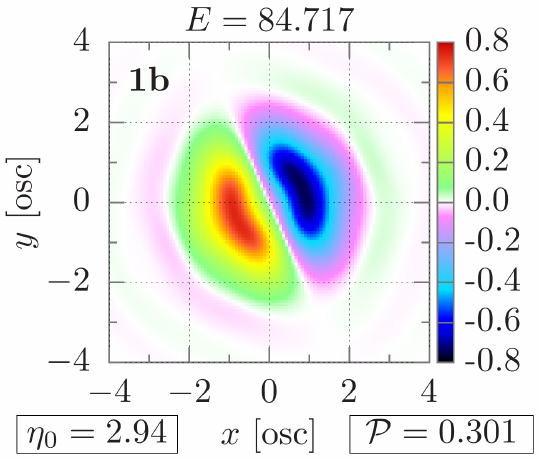}
\end{minipage}
\vspace{0.5\baselineskip}
\begin{minipage}{0.131\linewidth}
\begin{center}
${N_{1/2}=15}$
\end{center}
\vspace{-1.5\baselineskip}
\rule{\linewidth}{0.5pt}\\[0.5em]
\includegraphics[width=\linewidth]{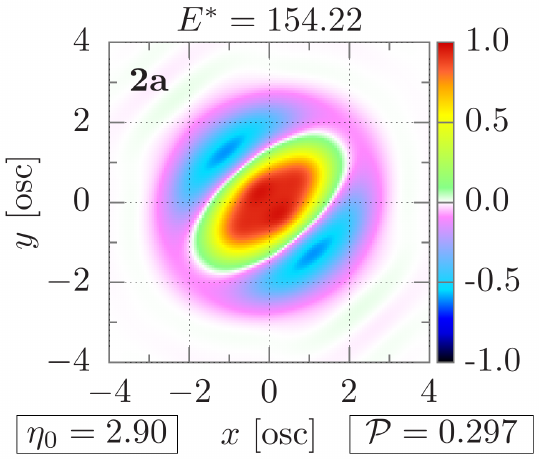}
\phantom{x}\vspace{-0.5em}
\includegraphics[width=\linewidth]{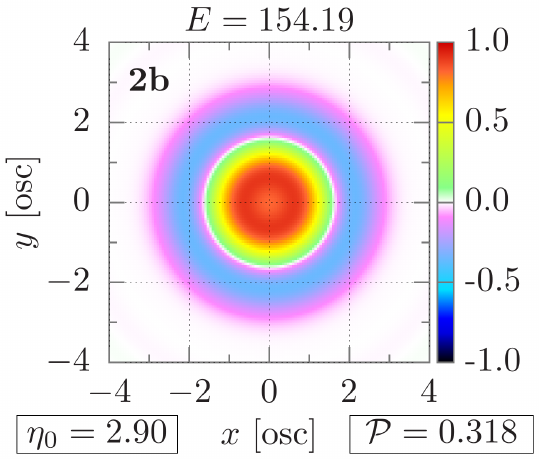}
\end{minipage}
\begin{minipage}{0.131\linewidth}
\begin{center}
${N_{1/2}=15}$
\end{center}
\vspace{-1.5\baselineskip}
\rule{\linewidth}{0.5pt}\\[0.5em]
\includegraphics[width=\linewidth]{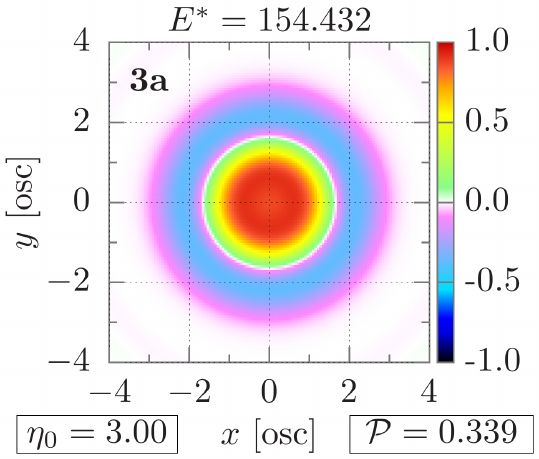}
\phantom{x}\vspace{-0.5em}
\includegraphics[width=\linewidth]{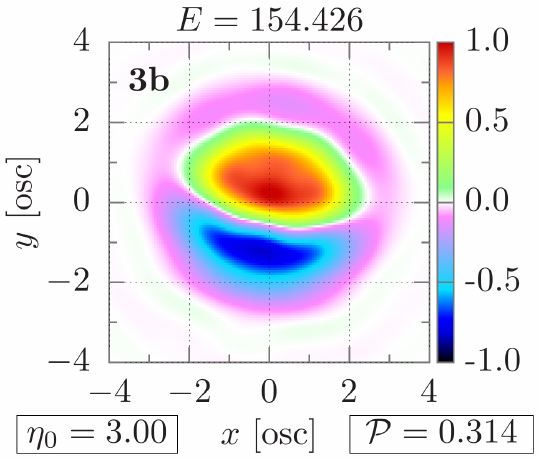}
\end{minipage}
\begin{minipage}{0.131\linewidth}
\begin{center}
${N_{1/2}=55}$
\end{center}
\vspace{-1.5\baselineskip}
\rule{\linewidth}{0.5pt}\\[0.5em]
\includegraphics[width=\linewidth]{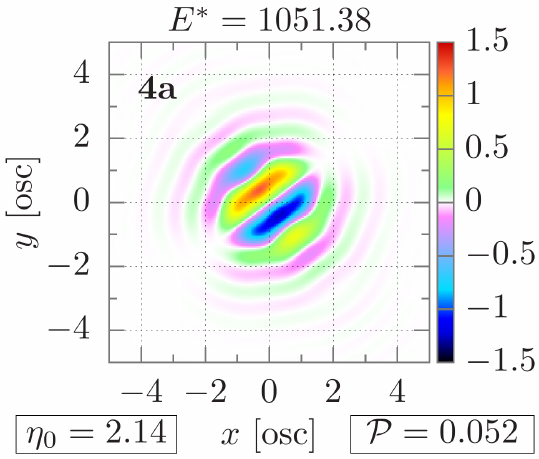}
\phantom{x}\vspace{-0.5em}
\includegraphics[width=\linewidth]{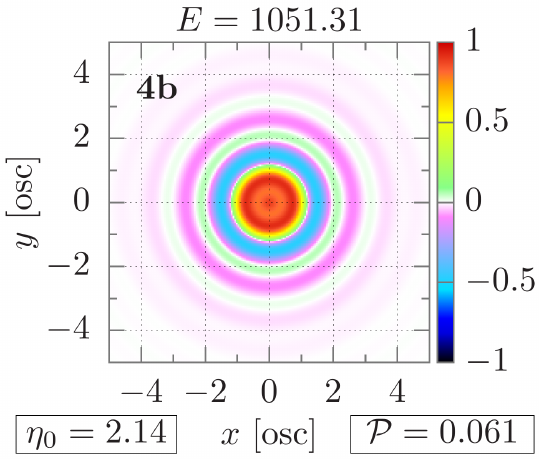}
\end{minipage}
\begin{minipage}{0.131\linewidth}
\begin{center}
${N_{1/2}=55}$
\end{center}
\vspace{-1.5\baselineskip}
\rule{\linewidth}{0.5pt}\\[0.5em]
\includegraphics[width=\linewidth]{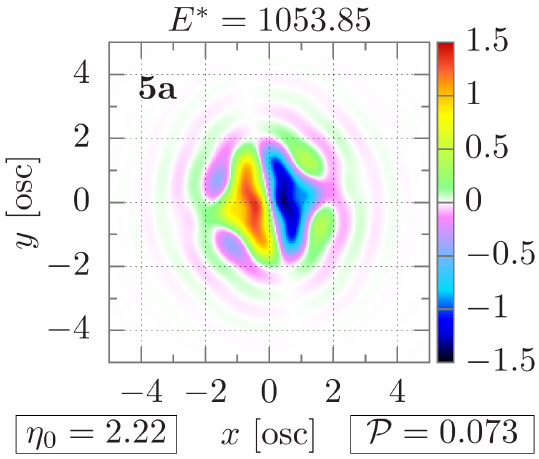}
\phantom{x}\vspace{-0.5em}
\includegraphics[width=\linewidth]{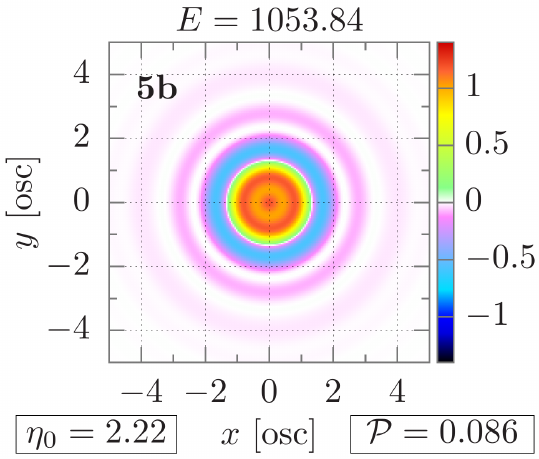}
\end{minipage}
\end{minipage}
\vspace{0.5\baselineskip}
\caption{\label{MetastableProfiles}Identification of metastable states. For ${N_{1/2}=10}$, $n_{3'}$ identifies ground-state separations that are isotropic, with energy $E_{\mathrm i}$, for ${2\lesssim\eta_0\lesssim3}$, and anisotropic, with energy $E_{\mathrm a}$ for ${\eta_0\gtrsim3}$ (left). An analogous observation holds for ${N_{1/2}=15}$ and ${N_{1/2}=55}$. The contour density plots identify candidates for metastable profiles (panels \textbf{1a-5a}; energies denoted by $E^*$) and ground-state densities (panels \textbf{1b-5b}) with energies $E$, marginally below $E^*$.}
\end{figure}

For the contact-type interactions studied in this work, the self-consistent DPFT loop of equation~(\ref{SCloop}) requires $\sim10^2$--$10^6$ iterations with admixing parameters ${\theta_s\sim0.2\mbox{--}0.6}$ for the densities to converge, in the sense of the correlation measure 
\begin{align}\label{SCcriterion}
\chi=1-\frac{2\,\vec{n}^{(i)}\cdot\vec n^{(i+1)}}{\big(\vec n^{(i)}\big)^2+\big(\vec n^{(i+1)}\big)^2} 
\end{align}
of the vectorized densities ${\vec{n}=\{n(\VEC r_1),\dots,n(\VEC r_G)\}}$ on the numerical grid of size $G$ to fall below a predefined threshold. We declare convergence once ${\chi<10^{-12}}$. The required number of iterations tends to be higher in the vicinity of $\eta_0^{\mathrm c}$, where the density profiles transition between qualitatively different phases within a narrow window of $\eta_0$. Depending on the interaction strength, we have encountered qualitatively different density profiles with relative energy differences of $\sim10^{-6}$--$10^{-3}$. Metastable profiles like those shown in figure~\ref{MetastableProfiles} (panels \textbf{1a}--\textbf{5a}) are therefore likely to be encountered in experiments alongside the actual ground-state profiles (panels \textbf{1b}--\textbf{5b}).

The self-consistent loop of equation~(\ref{SCloop}) can converge into different local minima. The following recipe increases the probability of converging to the actual ground-state densities---in our case, the global minimizers of the ($3'$-approximated) total energy functional---rather than to a metastable state. We always superimpose white noise\footnote{Without initial noise, the self-consistent loop may iterate isotropic densities in an isotropic trap (even if the global minimum features anisotropic densities) until accumulated rounding-off errors due to finite machine precision permit the approach towards the global minimum after many iterations.} with relative magnitude of $10\%$ on the initial densities $\VEC n^{(0)}$ to ensure that anisotropic density profiles are probed during the early stages of the self-consistent loop. Together with the sampling of the whole range of possible interaction strengths, this allows us to identify phase transitions between isotropic and anisotropic ground-state densities. In order to sample a sufficiently large space of densities, we impose noise on $\VEC n^{(i)}$ throughout the self-consistent loop akin to simulated annealing, with the noise magnitude incrementally decaying towards machine precision at the end of the loop. We converge to nearly machine precision in order to phase out intermediate metastable states on the way to the global energy minimum. But even with these measures in place, equation~(\ref{SCloop}) occasionally gets trapped in local energy minima, such that repeated runs are called for, especially near transitions between qualitatively different phases. Although the sequence of ground-state separation patterns for ${N_{1/2}=15}$ in figure~\ref{RenConDensityDiffn3p} suggests a sharp transition (around ${\eta_0=3}$) from isotropic separations towards the complete (symmetric) split into two semi-disks, panel \textbf{3b} of figure~\ref{MetastableProfiles} emerges smoothly from the metastable domain pattern encountered at lower $\eta_0$ (panel \textbf{2a}). We make an analogous observation for ${N_{1/2}=55}$, where the pattern with ${\eta_0=2.26}$ in figure~\ref{RenConDensityDiffn3p} emerges smoothly from the metastable branch, exemplified by the panels \textbf{4a} and \textbf{5a}.

We note that labeling density profiles as metastable is a challenging enterprise if based on (an approximate) energy evaluation alone. The minute energy differences in concert with numerical uncertainty are one reason: Differing density patterns and associated energies may emerge from different accuracy criteria for the numerics that have the potential to confuse the energy-based identification of metastability. Therefore, a high accuracy of internal computations and a high spatial resolution is required. But, more importantly, an unambiguous statement on metastability based on energy alone is possible only if the employed method (e.g., the energy functional) is exact and if the errors of the numerical procedures that target the ground state are negligible. Then, a density profile with higher energy is not the ground state density. Rigorously identifying the ground-state density is off the table in all other settings---in particular, if the energy functional is approximate. That said, making sure that numerical errors are insignificant and taking the approximate energy functional as a given, we can and do use energy as a criterion that decides on metastability if the corresponding DPFT calculations are initialized with identical input (except for the initial noise, which phases out after many loop iterations).

The multi-particle ground state for a rotationally symmetric trap is isotropic. We thus argue that the symmetry-broken density profiles, as obtained from our approximate HF and DFT schemes, represent some of the information encoded in the correlation functions (for example, density-density-correlations), which can be anisotropic also for a rotationally invariant many-body Hamiltonian with interactions~\cite{Perdew2021}. It is then conceivable that the actual isotropic ground-state can be constructed as an appropriate superposition of anisotropic states, each of which giving rise to `single-shot' density profiles akin to the approximate HF and DFT outcomes. However, such a superposition presents a fine-tuning problem that is irrelevant for both experiments and simulations---unless the isotropic ground-state density is robust against anisotropic perturbations that inevitably emerge both in the laboratory and in simulations. We have the latter situation, for example, if $\eta_0$ is well below $\eta_0^{\mathrm c}$. In our DPFT scheme, different types and magnitudes of perturbations on nonequilibrated densities during the self-consistent loop of equation~(\ref{SCloop}) determine the minimizers (equilibrated density patterns) at the (local or global) energy minima. For instance, initializing equation~(\ref{SCloop}) with an isotropic density and omitting noise, we observe convergence to an isotropic local minimizer and miss the anisotropic global minimizer because the local minimum is robust enough against numerical rounding-off errors.

\FloatBarrier

\section{Conclusions}\label{conclusions}

We mapped the para- to ferromagnetic phase transition of repulsive two-component Fermi gases in two dimensions beyond the local density approximation. By recovering the essential features of Hartree--Fock (HF) densities from density-potential functional theory (DPFT), supplied with systematically quantum-corrected semiclassical expressions for the particle densities that become more accurate for larger particle numbers, we gained quantitatively reliable density profiles that experimenters can expect to observe in realistic settings. Mapping both ground-state and metastable density configurations across interaction strengths for up to 10000 fermions, we predicted that strong contact-type interactions segregate the two fermion species into two semi-disks. We also found that the overlap of both species in this ferromagnetic phase can be reduced by increasing the particle number, which will likely suppress dimer formation.

We revealed several universal features of this phase transition across system sizes. All curves obtained from integrating the local polarizations ${|n_1-n_2|}$ essentially collapse onto a single graph and summarize the deviation from the paramagnetic state as a function of the contact-interaction strength. This confirms the long-standing prediction of a Stoner-type polarization behavior across particle numbers in terms of a universal gas parameter $\eta_0$, but we found the (nonuniform) harmonic trap responsible for stark quantitative differences to Quantum-Monte Carlo results for a uniform setting. Apart from the successful benchmarking against HF results, we gained confidence in the reliability of the DPFT results by observing that the density profiles transit smoothly from the para- to the ferromagnetic phase when increasing $\eta_0$ incrementally. In a sense, this observation also holds as $\eta_0$ crosses a critical value ${2.0\lesssim\eta_0^{\mathrm{c}}\lesssim3.1}$. Then, the anisotropic phase separations, which are metastable for ${\eta_0\lesssim\eta_0^{\mathrm{c}}}$, smoothly transform into the anisotropic ground-state phase, which shows isotropic separations for ${\eta_0\lesssim\eta_0^{\mathrm{c}}}$. With increasing particle number, we observe more and more qualitatively different intricate patterns of partially separated profiles, which are robust against small finite temperature, see appendix~\ref{AppendixDPFT}. They typically have relative energy differences of the order of $10^{-4}$ or less, which makes them likely to be observed in experiment in lieu of the actual ground state. In this regime of intermediate interaction strength, our markedly differing results from employing (i) the mean-field and (ii) the renormalized contact interaction, point in favor of the latter.

We also demonstrated the need to improve upon the Thomas--Fermi (TF) approximation for the kinetic energy, which fails to provide even a qualitatively reliable picture of the phase transition. We suspect the TF model's poor performance to originate in the inter-species contact-type interactions considered in this work. It is conceivable, however, that large systems with nonlocal interactions may be adequately addressed with the TF model. In any case, systematic corrections to the TF approximations have to either validate or replace the TF density expression. The semiclassical DPFT framework presented here is uniquely qualified to execute this task. Our DPFT code is part of the C++ software package `mpDPFT', available at \href{https://doi.org/10.5281/zenodo.4774448}{https://doi.org/10.5281/zenodo.4774448}. Imbalanced mixtures can be addressed without any modifications of the code, and implementations of additional density-dependent interaction functionals are straightforward. DPFT is a multi-purpose tool for addressing quantum-many-body problems in one-, two-, and three-dimensional geometries, such as multi-component dipolar Fermi gases. They are recently coming into focus with first experimental realizations \cite{Baier2018,DeMarco2019,Neri2020} and are candidates for showing even richer phase transitions than the ones predicted here for contact-type interactions.

\acknowledgments
We are grateful to Berthold-Georg Englert for valuable insights and feedback. J.~H.~H. acknowledges the financial support of the Graduate School for Integrative Science \& Engineering at the National University of Singapore. This work is partially funded by the Singapore Ministry of Education and the National Research Foundation of Singapore.
P.~T.~G. is financed from the (Polish) National Science Center Grants 2018/29/B/ST2/01308 and 2020/36/T/ST2/00065. 
K.~Rz. is supported from the (Polish) National Science Center Grant 2018/29/B/ST2/01308.
The Center for Theoretical Physics of the Polish Academy of Sciences is a member of the National Laboratory of Atomic, Molecular and Optical Physics (KL FAMO). Part of the results were obtained using computers of the Computer Center of the University of Bia{\l}ystok.

\appendix

\section{\label{AppendixDPFT}Details on DPFT densities and associated kinetic energies}

\textbf{`Airy-averaged' densities.} As a potential alternative to $n_{3'}$, we consider the `Airy-averaged' densities $n_{\mathrm{Ai}}^T$ from a second DPFT approximation scheme, which is worked out in references~\cite{Englert1988,Trappe2016a,Trappe2017} and is based on expressing the trace in equation~(\ref{tracef}) as a classical phase-space integral that is systematically approximated with the help of Airy functions $\mathrm{Ai}(\,)$. Manuscripts that cover the 1D \cite{Trappe2021a} and 3D \cite{Trappe2021b} formulae are in preparation. Here, we focus on the 2D situation, which is covered extensively in references~\cite{Trappe2016a,Trappe2017}.

The Airy-averaged 2D ground-state densities ${n_{\mathrm{Ai}}^{T=0}}$ derived in reference~\cite{Trappe2017} exhibit unphysical oscillations in the vicinity of positions~$\VEC r$ where $\nab V(\VEC r)=0$. By introducing a small but finite temperature $T$, we obtain the (2D) density expression
\begin{align}\label{nAirygr0}
n_{\mathrm{Ai}}^T(\VEC r)=\int\d x\,\mathrm{Ai}(x)&\left\{\frac{k_{\mathrm{B}}T}{\pi\,\mathcal{U}}\mathrm{log}\left(1+\frac{1}{\zeta}\right)-\frac{\nab^2V}{12\pi k_{\mathrm{B}}T}\frac{\zeta}{(1+\zeta)^2}\right\},
\end{align}
which is well-behaved everywhere \cite{Trappe2017}; ${\zeta(x,\VEC r,T)=\exp\{\left[V(\VEC r)-\mu-x\,a(\VEC r)\right]/k_{\mathrm{B}}T\}}$ and ${a(\VEC r)=\frac{\mathcal{U}^{1/3}}{2}|\nab V|^{2/3}}$.\\

\begin{figure}[htbp!]
\begin{flushleft}
\includegraphics[width=0.37\linewidth]{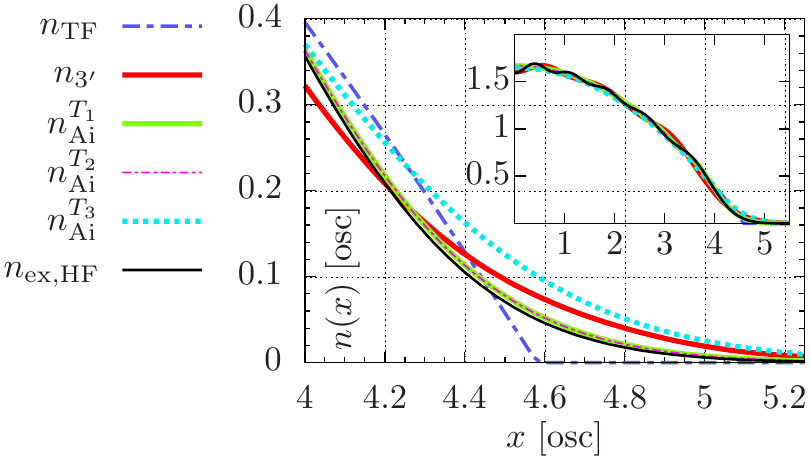}
\hspace{1em}
\includegraphics[width=0.30\linewidth]{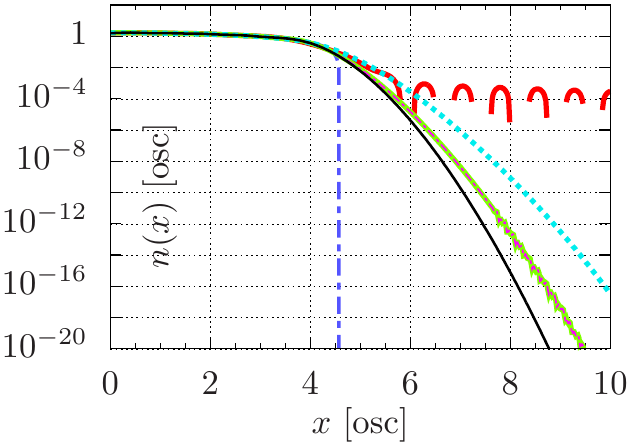}
\hspace{1em}
\includegraphics[width=0.265\linewidth]{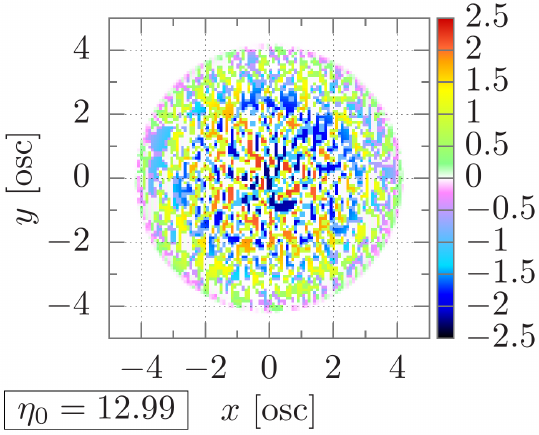}
\end{flushleft}
\caption{\label{TFvsn3p} We illustrate the accuracy of our semiclassical particle densities and the inadequacy of the Thomas--Fermi approximation for the two-component mixture. Left: Compared with the exact (isotropic) density $n_{\mathrm{ex}}$ of ${N=55}$ noninteracting fully polarized fermions, the quasiclassical TF-approximated density is reasonably adequate, although the missing decay into the classically forbidden region is only recovered with quantum-corrected density formulae like $n_{3'}$ and $n_{\mathrm{Ai}}^T$ (main plot; harmonic oscillator units [osc]). The bulk of the semiclassical density profiles (inset) and the total energies (${E_{\mathrm{TF}}\approx769.13\,\hbar\omega}$, ${E_{3'}\approx774.39\,\hbar\omega}$, ${E_{\mathrm{Ai}}^{T_1}\approx E_{\mathrm{Ai}}^{T_2}\approx770.61\,\hbar\omega}$, ${E_{\mathrm{Ai}}^{T_3}\approx770.36\,\hbar\omega}$) are close to the exact quantities, as expected for over one hundred particles. The Friedel oscillations in the bulk of $n_{\mathrm{ex}}$ are not reproduced by our semiclassical densities, which rather give an approximate average account of the exact oscillations. Center: The densities $n_{\mathrm{Ai}}^{T_{1/2}}$ approximate $n_{\mathrm{ex}}$ well over $\sim20$ orders of magnitude, while $n_{3'}$ exhibits unphysical Bessel-function-induced oscillations far into the classically forbidden region. This imperfection of $n_{3'}$ is of minor concern for the present work since the phase separations between the two fermion species develop in the bulk. Right: For the two-component mixture at renormalized contact interaction strength ${\eta_0=12.99}$, the TF approximation of ${n_1-n_2}$ simply amplifies the noise imposed at the start of the self-consistent loop of equation~(\ref{SCloop}) because different positions are decoupled for contact-type interactions (see section~\ref{SectionTF}, and~\cite{Trappe2016} for the 3D case of the mean-field contact interaction). This shortcoming of TF is inherited by $n_{\mathrm{Ai}}^T$ and contrasts with the performance of the nonlocal $n_{3'}$ approximation, whose converged smooth density profiles (for ${\eta_0=12.99}$) roughly envelop the `TF noise' depicted here, cf.~figure~\ref{RenConDensityDiffn3p}.}
\end{figure}

\textbf{Benchmarking of DPFT densities against exact results.} In figure~\ref{TFvsn3p} we benchmark the semiclassical density formulae $n_{3'}$ and $n_{\mathrm{Ai}}^T$ against exact densities of a harmonically trapped Fermi gas in 2D. For $n_{\mathrm{Ai}}^T$, we use temperatures $T_i$ such that $k_{\mathrm{B}}T_1=10^{-6}\,\hbar\omega$, $k_{\mathrm{B}}T_2=10^{-6}E_{\mathrm{ex}}$ (with the exact energy ${E_{\mathrm{ex}}=770\,\hbar\omega}$ of the two-component system of ${N=N_1+N_2=55+55}$ particles), and $k_{\mathrm{B}}T_3=\hbar\omega$. Both $n_{\mathrm{Ai}}^{T_1}$ and $n_{\mathrm{Ai}}^{T_2}$ are close to $n_{\mathrm{ex}}$ and showcase the capacity of semiclassical DPFT for describing the region around the quantum-classical boundary. We also see that $T_2$ is small enough for extracting ground-state properties from the finite-temperature formula $n_{\mathrm{Ai}}^{T}$, see equation~(\ref{nAirygr0}) in appendix~\ref{AppendixDPFT}. However, as argued in the following, $n_{\mathrm{Ai}}^T$ is not suitable to address multi-component systems with contact-type interactions due to the bulk properties of $n_{\mathrm{Ai}}^T$ inherited from the TF density---despite $n_{\mathrm{Ai}}^T$ being generically more accurate than $n_{3'}$.

The leading term in equation~(\ref{nAirygr0}) with the natural logarithm $\mathrm{log}(\,)$ recovers the (finite-temperature) TF density for uniform effective potentials (i.e., ${\nab V=0}$ everywhere). Equation (\ref{nAirygr0}) is exact up to the leading gradient correction $\big(\mathcal O(\nab^2)\big)$, and thus presents a systematic correction to the TF approximation $\big(\mathcal O(\nab^0)\big)$. The `Airy-average' in equation~(\ref{nAirygr0}) also contains higher-order gradient corrections that provide an accurate density tail across the boundary of classically allowed and forbidden regions of $V$, where the TF approximation can fail epically, even if supplemented with the leading gradient correction \cite{Trappe2017}. However, in this work we focus on the bulk, not on the evanescent tails. The two fermion components separate in the bulk, where $n_{\mathrm{TF}}^T$ is the dominant component of (and transfers its inadequacy to) $n_{\mathrm{Ai}}^T$. Indeed, we find that the semilocal nature of $n_{\mathrm{Ai}}^T$, which stems from the derivatives of $V$, does not prevent the convergence into random domains of partial polarization, very similar to figure~\ref{TFvsn3p} (right). In contrast, $n_{3'}$ is fully nonlocal and retains less of the TF characteristics in the bulk. Both these features promote smooth density profiles when using $n_{3'}$ even for large particle numbers.\\

\textbf{Numerical details on $\mathbf{n_{3'}}$.} The approximate nature of $n_{3'}$ can lead to locally negative densities (exhibited by the `gaps' in the evanescent region of the graph in the central panel of figure~\ref{TFvsn3p}), although this effect can be regarded negligible for $\gtrsim10$ particles. Since the renormalized contact interaction requires strictly positive densities everywhere, cf.~equation~(\ref{eta}), we replace $n_{3'}$ by ${[n_{3'}]_++10^{-16}}$ when evaluating equations~(\ref{epsilonint}) and (\ref{epsilonintderiv}). We determine the chemical potentials $\mu_s$ in each iteration of the DPFT loop via an adaptive bisection algorithm, enforcing a relative accuracy of at least $10^{-6}$ for the particle numbers $N_s$.\\

\textbf{$\mathbf{n_{3'}}$ at finite temperature.} The derivation of the finite-temperature version
\begin{align}\label{n3pT}
n_{3'}^T(\VEC r)&=\frac{g}{\Gamma(D/2)}\left(\frac{k_{\mathrm{B}}T}{2\pi\mathcal{U}}\right)^{D/2}\int_0^\infty\d y\,\mathcal{F}^{-1}\left\{\mathcal{F}\left\{f_y(\VEC r')\right\}(\VEC k)\,g_y^D(k)\right\}(\VEC r)
\end{align}
of $n_{3'}(\VEC r)$ will be given elsewhere \cite{Trappe2021b}. Here, $\Gamma(\,)$ denotes the Gamma function,
\begin{align}\label{fy}
f_y(\VEC r')=\exp\left\{\tau\big(1-V(\VEC r')/\mu\big)-y\,\exp\left[\tau\big(1-V(\VEC r')/\mu\big)\right]\right\},
\end{align}
with ${\tau=\mu/(k_{\mathrm{B}}T)}$, and
\begin{align}\label{gyD}
g_y^D(k)=\int_0^\infty\d x\, x^{D/2-1}\,\exp\left[-y\,\exp\left(x+\kappa\right)\right]
\end{align}
is easily tabulated for all required values of ${\kappa=(\hbar k)^2/(8m\,k_{\mathrm{B}}T)=\mathcal{U}\,k^2/(k_{\mathrm{B}}T)}$, where $k$ is the magnitude of the wave vector $\VEC k$. The computational cost of $n_{3'}^T(\VEC r)$ scales like $G\,\log G$. This contrasts with equation~(\ref{n3p}), where the density $n_{3'}(\VEC r)$ at each of the $G$ grid points, indexed by $\VEC r$, requires a summation over the whole grid. Naturally, there is a trade-off between grid size and accurate enough evaluation of the $y$-integral---as a rule of thumb, $n_{3'}^T$ outperforms $n_{3'}$ for ${G\gtrsim 50^3\approx350^2}$ and $n_{3'}^{\mathcal F}$ for ${G\gtrsim 100^3=1000^2}$. That is, the computational efficiency of $n_{3'}^T$ exceeds that of $n_{3'}^{\mathcal F}$ for most 3D applications. In 2D settings, however, it is expedient to use $n_{3'}^{\mathcal F}$ instead of $n_{3'}^T$, unless very high spatial resolution is required or if finite-temperature observables are targeted.

\begin{figure}[htb!]
\includegraphics[height=0.125\linewidth]{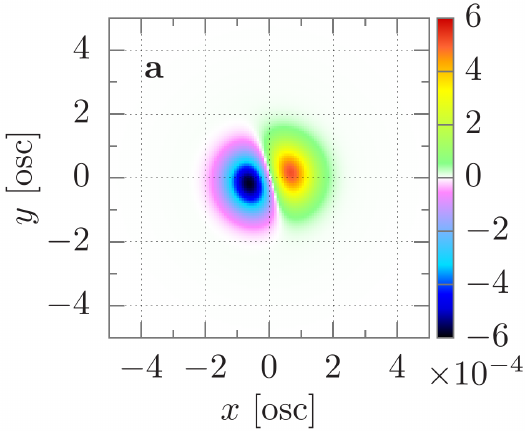}
\includegraphics[height=0.125\linewidth]{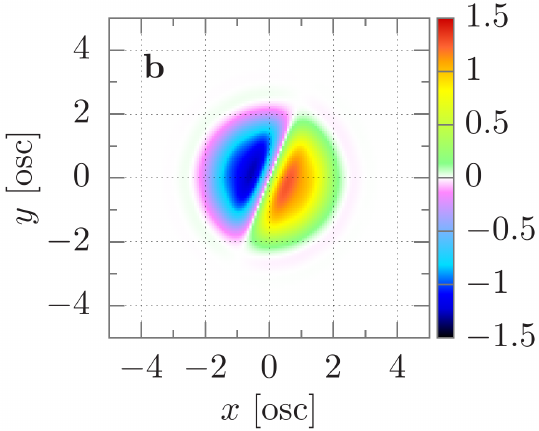}
\includegraphics[height=0.125\linewidth]{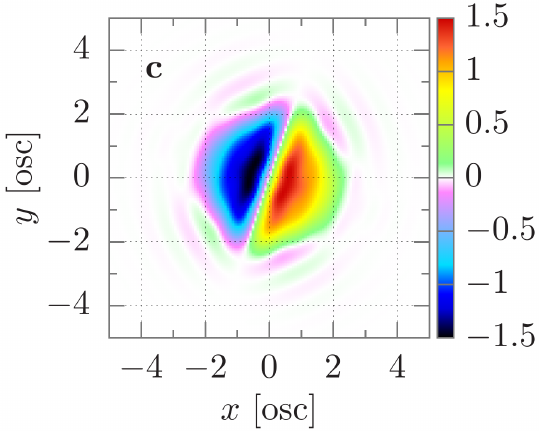}
\includegraphics[height=0.125\linewidth]{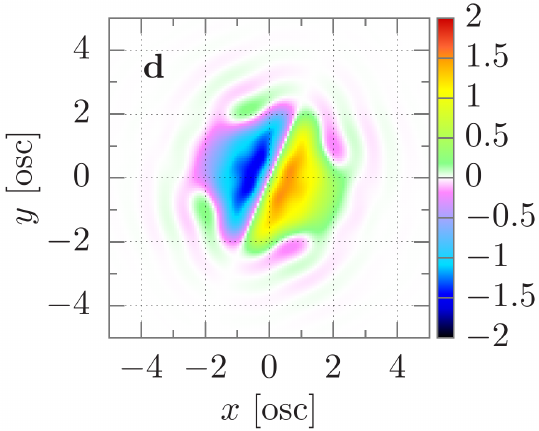}
\includegraphics[height=0.125\linewidth]{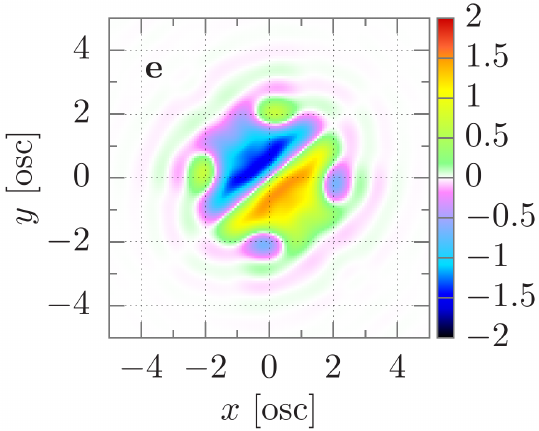}
\includegraphics[height=0.125\linewidth]{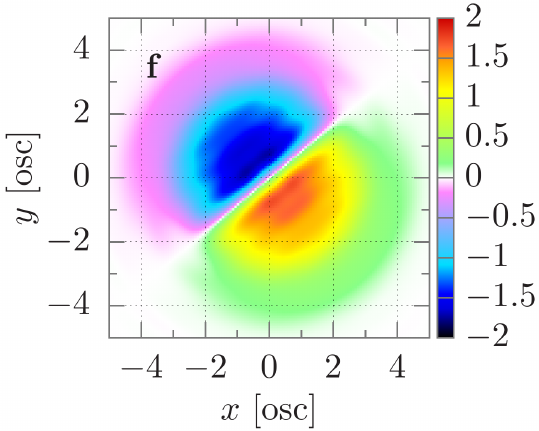}
\caption{\label{CompareFiniteT}In view of experiments conducted at finite temperature, we illustrate at the example of $N_s=55$ with ${\eta_0=2.35}$ that the structure of the density profiles is robust against small nonzero temperatures. Here, we use ${k_{\mathrm{B}}T=2,1,0.2,0.1,0}$ (in [osc] units) for the panels~\textbf{a}--\textbf{e}, and $n_{3'}^T$ ($n_{3'}^{\mathcal{F}}$) for finite (zero) temperature. Panel~\textbf{f} depicts the converged density difference after adding the Coulomb interaction of equation~(\ref{Coulomb}) with ${\gamma=1/20}$ to the renormalized contact interaction. We transit from panel~\textbf{f} to panel~\textbf{e}, thereby recovering the ground-state configuration, by gradually reducing $\gamma$ to zero over thousands of iterations.}
\end{figure}

Our results in figure~\ref{CompareFiniteT} demonstrate that $n_{3'}^T$ transforms smoothly into the ground-state density $n_{3'}$ as $T$ tends to zero. We thus expect that the intricate phase separations at intermediate interaction strengths can also be observed in experiments performed at small but finite temperatures.\\

\FloatBarrier

\textbf{Regularization with a transient Coulomb interaction.} For ${N_{1/2}=5000}$ in the ferromagnetic phase, we arrive at the ground-state configuration that splits the Fermi components symmetrically by initializing the self-consistent DPFT loop with a symmetrically split configuration. We aim at a smooth DPFT equilibration into the actual ground-state profiles by counteracting the tendency of large particle numbers at strong contact-type interactions to simply enhance initially imposed noise. While this could in principle be achieved with the finite-temperature formula $n_{3'}^T$, which washes out any noise and oscillatory structures if $T$ is high enough, see figure~\ref{CompareFiniteT}, $n_{3'}^T$ incurs high computational cost as $T$ comes close to zero. We therefore regularize by adding an artificial mutually repulsive long-range (viz., nonlocal) interaction to the effective potential $V_s$, specifically, the Coulomb interaction potential
\begin{align}\label{Coulomb}
V_s^{\mathrm{Coul}}[n_{s'}](\VEC r)=\gamma\,\big(1-\delta_{ss'}\big)\int(\d\VEC r')\,\frac{n_{s'}(\VEC r')}{|\VEC r-\VEC r'|}
\end{align}
in 2D. We express $V_s^{\mathrm{Coul}}$ in terms of Fourier transforms for an efficient numerical implementation and regularize the Coulomb singularity in reciprocal space that is introduced through the numerical discretization. In the case of ${N_{1/2}=5000}$, we employ ${\gamma=1/4}$ in harmonic oscillator units.\\

\textbf{Kinetic energy in $\mathbf{U_{3'}}$-approximation.} We obtain approximations of the (ground-state) kinetic energy
\begin{align}
E_{\mathrm{kin}}=-\frac{\hbar^2}{2m}\int(\d\VEC r)\,\left(\nab_{\VEC r}^2\, n^{(1)}(\VEC r;\VEC r')\right)_{\VEC r'=\VEC r}
\end{align}
by deriving approximate one-body reduced density matrices ${n^{(1)}(\VEC r;\VEC r')}$ in terms of the effective potential $V$. With ${H=T+V}$, where ${T=\VEC P^2/(2m)}$ is the single-particle kinetic-energy operator, equation~(\ref{LegendreTF}) yields\footnote{In detail, $E_1[V-\mu]-\int(\d\VEC r)\,\big(V(\VEC r)-\mu\big)\,n(\VEC r)=\tr\{(T+V-\mu)\,\Theta(\mu-H)\}+\int(\d\VEC r)\,\big(\mu-V(\VEC r)\big)\,n(\VEC r)=\tr\{T\,\Theta(\mu-H)\}=g\int(\d\VEC r)(\d\VEC r')\,\bok{\VEC r}{\left(-\frac{\hbar^2}{2m}\nab_{\VEC r}^2\right)}{\VEC r'}\bok{\VEC r'}{\Theta(\mu-H)}{\VEC r}=g\int(\d\VEC r)(\d\VEC r')\,\delta(\VEC r-\VEC r')\,\left(-\frac{\hbar^2}{2m}\nab_{\VEC r}^2\right)\bok{\VEC r'}{\Theta(\mu-H)}{\VEC r}$.}
\begin{align}
E_{\mathrm{kin}}=-g\frac{\hbar^2}{2m}\int(\d\VEC r)\,\left(\nab_{\VEC r}^2\bok{\VEC r}{\Theta(\mu-H)}{\VEC r'}\right)_{\VEC r'=\VEC r}.
\end{align}
In the spirit of equation~(\ref{nSTA}), we thus identify the approximate one-body reduced density matrix
\begin{align}\label{1RDM3p}
n^{(1)}_{3'}(\VEC r;\VEC r')&=g\Int\frac{\d t}{2\pi\I t}\,\e{\frac{\I t}{\hbar}\mu}\,\bok{\VEC r}{U_{3'}(t)}{\VEC r'}=g\int(\d\VEC r'')\left(\frac{k_{3'}}{2\pi b}\right)^D J_D(2b\,k_{3'}),
\end{align}
which is consistent with the Suzuki--Trotter approximation inherent to $n_{3'}$. Equation~(\ref{1RDM3p}) follows the structure of equation~(\ref{n3p}) with ${b=\sqrt{r''^2+(\VEC r''+\VEC r-\VEC r')^2}}$ and yields equation~(\ref{Ekin3p}):
\begin{align}
E_{\mathrm{kin}}^{(3')}&=-\frac{\hbar^2}{2m}\int(\d\VEC r)\,\left[\nab_{\VEC r}^2 n^{(1)}_{3'}(\VEC r;\VEC r')\right]_{\VEC r'=\VEC r}=\frac{g\,\Omega_D}{(2\pi\hbar)^D\,(2D+4)\, m}\int(\d\VEC r)\,\big[2m\big(\mu-V(\VEC r)\big)\big]_+^{\frac{D+2}{2}},
\end{align} 
which can also be used in lieu of the finite-temperature kinetic energy $E_{\mathrm{kin}}^{(3'),T}$ for small enough $T$.\\

\section{\label{AppendixHF}Derivation of the Hartree--Fock equations}

In the single-particle approach each fermion is described by a spinor $\phi_i(j)$ built from spin-orbitals 
$\varphi_i(\VEC r_j,s_j)$, where $\VEC r$ and $s$ are spatial 
 and spin coordinates, respectively. The indices $i$ and $j$ take values from $1$ to
 $N$, where $N$ is the total number of fermions. The spin-orbitals obey
\begin{align}
  \sum_{s=1}^{S}\int (\d\VEC r)\, \varphi_i^{*}(\VEC r,s)\,
   \varphi_j(\VEC r,s)
= \delta_{ij},
\label{OrthoNorm}
\end{align}
where $S$ is the number of spin components. For noninteracting fermions the 
solution of the multi-particle Schr\"{o}dinger equation is
\begin{align}
 \Psi = \frac{1}{\sqrt{N!}} \sum_{P} \mathrm{sgn}(P)\, \phi_{P_1}(1)\, \phi_{P_2}(2) \cdot ... \cdot \phi_{P_N}(N),
 \label{AntAnz}
\end{align}
where $P$ is a permutation of $N$ elements and each $\phi_i(j)$ is the solution 
of the single-particle Schr\"{o}dinger equation. The lowest-lying set of solutions has to be taken for the ground state. The total one-particle density is given by ${n(\VEC r) = \sum_{s=1}^S\, n_s(\VEC r)}$, where ${n_s(\VEC r) = \sum_{i=1}^N\,|\varphi_i(\VEC r,s)|^2}$ is the one-particle density of the spin component $s$.

When the interaction is turned on, 
Equation~(\ref{AntAnz}) can be taken as a variational ansatz.
The ground state of a system is found by minimizing the total energy functional with respect to $\{\varphi_i^*\}$ and $\{\varphi_j\}$. This functional is given by
\begin{align}\label{EHF}
 E[\varphi_1^*,...,\varphi_N^*,\varphi_1,...,\varphi_N] \equiv 
 E[\{\varphi_i^*\},\{\varphi_j\}] =
 \sum_{s_1 , ... , s_N = 1}^S \int (\d\VEC r_1 ... \d\VEC r_N)\, \Psi^* \hat{H}\, \Psi,
\end{align}
where
\begin{align}
 \hat{H} = \sum_{i=1}^N\left[-\frac{\hbar^2}{2m}\nabla_i^2+V_{\mathrm{ext}}(\VEC r_i)\right]+
 \sum_{i<j} V_{\mathrm{int}}(\VEC r_i - \VEC r_j)
\end{align}
is the Hamiltonian of our system.
Equation~(\ref{EHF}) can be rewritten as
\begin{align}
 E[\{\varphi_i^*\},\{\varphi_j\}] = \sum_{i=1}^N h_i + 
 \sum_{i<j}\left(K_{ij}-J_{ij}\right),
 \label{HFEnergy}
\end{align}
where $h_i$ is the average one-particle energy, $K_{ij}$ is the average interaction
energy between the states $\varphi_i$ and $\varphi_j$, and $J_{ij}$ is the interchange energy:
\begin{eqnarray}
 h_i & = & \sum_{s=1}^S \int (\d\VEC r)\, \varphi_i^{*}(\VEC r,s)
 \left[-\frac{\hbar^2}{2m}\nabla^2+V_{\mathrm{ext}}(\VEC r)\right] \varphi_i(\VEC r,s)\, ,\\
 K_{ij} & = & \sum_{s,s^{\prime}=1}^S \int (\d\VEC r)(\d\VEC r^{\prime})\, 
 |\varphi_i(\VEC r,s)|^2 \, V_{\mathrm{int}}^{ss^{\prime}}(\VEC r - \VEC r^{\prime}) \, 
 |\varphi_j(\VEC r^{\prime},s^{\prime})|^2\, ,\\
 J_{ij} & = & \sum_{s,s^{\prime}=1}^S \int (\d\VEC r)(\d\VEC r^{\prime})\,
 \varphi_i^{*}(\VEC r,s)\, \varphi_j^{*}(\VEC r^{\prime},s^{\prime})\,
 V_{\mathrm{int}}^{ss^{\prime}}(\VEC r - \VEC r^{\prime}) \,
 \varphi_i(\VEC r^{\prime},s^{\prime})\, \varphi_j(\VEC r,s).
\end{eqnarray}
We ensure that the spin-orbitals obey equation~(\ref{OrthoNorm}) by supplementing equation~(\ref{HFEnergy}) with Lagrange multipliers $\lambda_{ij}$,
\begin{align}
 F[\{\varphi_i^*\},\{\varphi_j\}] = E[\{\varphi_i^*\},\{\varphi_j\}] -
 \sum_{i,j=1}^N \lambda_{ij}
 \left(\sum_{s=1}^S\int (\d\VEC r)\, \varphi_i^{*}(\VEC r,s)\,
 \varphi_j(\VEC r,s) - \delta_{ij}\right).
\end{align}
The global minimum of $E[\{\varphi_i^*\},\{\varphi_j\}]$ is obtained from the vanishing total variation
\begin{align}
 \delta F = \sum_{i=1}^N \frac{\partial F}{\partial \varphi_i^{*}} \,
 \delta \varphi_i^{*} +
 \sum_{i=1}^N \frac{\partial F}{\partial \varphi_i} \,
 \delta \varphi_i = 0
 \label{Variation}
\end{align}
of $F[\{\varphi_i^*\},\{\varphi_j\}]$. Equation~(\ref{Variation}) is fulfilled for any $\delta \varphi_i^{*}$ and
$\delta \varphi_i$ if all partial derivatives vanish,
\begin{eqnarray}
 \frac{\partial F}{\partial \varphi_i^{*}} & = & 0,\label{Var1}\\
 \frac{\partial F}{\partial \varphi_i} & = & 0.
 \label{Var2}
\end{eqnarray}
The set of equations~(\ref{Var1}) for all ${i=1,...,N}$ gives the Hartree--Fock equations for $\varphi_i$. The second set from equations~(\ref{Var2}) is complex conjugate version
of these equations. For the first set, one obtains
\begin{align}
 \left[ \hat{h}(\VEC r) + \sum_{j=1}^N\hat{K}_j(\VEC r) - 
 \sum_{j=1}^N\hat{J}_j(\VEC r) \right] \varphi_i(\VEC r, s) = \varepsilon_i^{(s)} \, \varphi_i(\VEC r, s),
 \label{HFeq}
\end{align}
where the operators $\hat{h}(\VEC r)$, $\hat{K}_j(\VEC r)$, and $\hat{J}_j(\VEC r)$ 
are defined through
\begin{eqnarray}
 \hat{h}(\VEC r) \, \varphi_i(\VEC r, s) & = &
 \left[-\frac{\hbar^2}{2m}\nabla^2+V_{\mathrm{ext}}(\VEC r)\right]
 \varphi_i(\VEC r, s),\\
 \hat{K}_j(\VEC r) \, \varphi_i(\VEC r, s) & = &
 \sum_{s^{\prime}=1}^S \int (\d\VEC r^{\prime})\,
 \varphi_j^{*}(\VEC r^{\prime}, s^{\prime}) 
 V_{\mathrm{int}}^{ss^{\prime}}(\VEC r - \VEC r^{\prime})\, \varphi_j(\VEC r^{\prime}, s^{\prime})\,
 \varphi_i(\VEC r, s),\\
 \hat{J}_j(\VEC r) \, \varphi_i(\VEC r, s) & = &
 \sum_{s^{\prime}=1}^S \int (\d\VEC r^{\prime})\, 
 \varphi_j^{*}(\VEC r^{\prime}, s^{\prime}) 
 V_{\mathrm{int}}^{ss^{\prime}}(\VEC r - \VEC r^{\prime})\, \varphi_i(\VEC r^{\prime}, s^{\prime})\,
 \varphi_j(\VEC r, s).
\end{eqnarray}
This set of equations allows us to find the optimal spin-orbitals. The Hartree--Fock energy $E_{\mathrm{HF}}$ is obtained by substituting 
these spin-orbitals into equation~(\ref{HFEnergy}).

Now we consider the two-component case, that is, ${s,s^{\prime}\in\{1,2\}}$. In each component 
we have $N/2$ particles. We assume that for the first half of particles (and vice versa for the second half) spin-down orbitals are unoccupied and spin-up orbitals are occupied, that is,
\begin{equation}
\left[
\begin{array}{c}
   \varphi_i(\VEC r,1) \\
   \varphi_i(\VEC r,2)
  \end{array}
  \right] = \left[
  \begin{array}{c}
   \varphi_i^{(1)}(\VEC r) \\
   0
\end{array}
\right]\mbox{ and }
\left[
  \begin{array}{c}
   \varphi_{N/2+i}(\VEC r,1) \\
   \varphi_{N/2+i}(\VEC r,2)
  \end{array}
  \right] = \left[
  \begin{array}{c}
  0\\
   \varphi_i^{(2)}(\VEC r)
  \end{array}
  \right],
\end{equation}
where ${i=1,...,N/2}$ (with $N$ even).
Moreover, we assume the mean-field contact interaction
\begin{equation}
   V_{\mathrm{int}}^{12}(\VEC r-\VEC r^{\prime}) = 
   \alpha\,\delta(\VEC r-\VEC r^{\prime})
\end{equation}
between components and no interaction within components, ${V_{\mathrm{int}}^{11}(\VEC r-\VEC r^{\prime}) = V_{\mathrm{int}}^{22}(\VEC r-\VEC r^{\prime}) = 0}$. Then, equation~(\ref{HFeq}) reads
\begin{align}\label{HFeqsimple}
 \left[ -\frac{\hbar^2}{2 m} \nabla^2 + V_{\mathrm{ext}}(\VEC r)
+ \alpha\, n_s(\VEC r) \; \right] \; \varphi_i^{(s')} (\VEC r) & = \varepsilon_i^{(s')} \, \varphi_i^{(s')}(\VEC r),
\end{align}
where ${s\not=s'}$ and ${i=1,...,N/2}$. Here, the one-particle densities ${n_s(\VEC r) = \sum_{i=1}^{N/2} |\varphi_i^{(s)} (\VEC r)|^2}$ of the spin components $s$ yield the total one-particle density ${n(\VEC r) = n_1(\VEC r)+n_2(\VEC r)}$. For studying renormalized interaction terms, we replace ${\alpha\, n_{1/2}(\VEC r)}$ in equation~(\ref{HFeqsimple}) by equation~(\ref{epsilonintderiv}).

\bibliography{library}

\end{document}